\documentclass[wcd, manuscript]{copernicus}

\usepackage{graphicx}
\usepackage{hyperref}
\usepackage{booktabs}

\begin{document}

\nolinenumbers

\title{Zonal asymmetries control the response of atmospheric blocking to Arctic warming in an aquaplanet experiment}



\Author[1,2][michele.filippucci@unitn.it]{Michele}{Filippucci} 
\Author[3]{Stephen}{I. Thomson}
\Author[3,4]{Neil}{Lewis}
\Author[1,5]{Simona}{Bordoni}

\affil[1]{Department of Civil, Environmental and Mechanical Engineering, University of Trento, Trento, Italy}
\affil[2]{Istituto Universitario Superiore di Pavia, Pavia, Italy}
\affil[3]{Department of Mathematics and Statistics, University of Exeter, Exeter, UK}
\affil[4]{Department of Physics and Astronomy, University of Exeter, Exeter, UK}
\affil[5]{Center Agriculture Food Environment, University of Trento, Trento, Italy}



\runningtitle{TEXT}

\runningauthor{TEXT}

\received{}
\pubdiscuss{} 
\revised{}
\accepted{}
\published{}


\firstpage{1}

\maketitle

\begin{abstract}

In recent years a weak but robust response of mean midlatitude circulation to Arctic amplification (AA) has emerged from modeling experiments. However, open questions remain about the mechanisms linking such circulation differences to weather extremes in the midlatitudes. In this study we investigate such mechanisms and the importance of zonal asymmetries in shaping the atmospheric blocking response to AA. We perform idealized aquaplanet simulations in two configurations: a zonally symmetric setup and a zonally asymmetric experiment featuring a localized midlatitude storm track. For each configuration, we examine the response to AA by imposing an anomalous surface heating in the polar region.

In the zonally symmetric configuration atmospheric blocking increases uniformly with AA from mid to high latitudes. In the asymmetric configuration, the response is more complex; instead of a zonally uniform response, we observe an upstream displacement of the blocking maximum, which sits at the exit of the localized storm track. We interpret these changes through the lens of the Traffic Jam theory by diagnosing the carrying capacity of the midlatitude flow. In both configurations, the zonally averaged increase in blocking is primarily driven by a weakening of the zonal winds, which reduces the Doppler-shifted Rossby wave group velocity and, in turn, decreases the flow carrying capacity. While the reduction in carrying capacity has similar characteristics in the two configurations, in the asymmetric case it leads to an upstream shift of blocking frequency as a direct consequence of the threshold behavior of blocking onset that lies at the core of the Traffic Jam theory. This mechanism, which has received limited attention so far, highlights the importance of mean circulation characteristics in shaping the blocking response to external forcing such as Arctic warming. 

Our results may therefore help to understand blocking responses to AA in more comprehensive models, by providing a physically consistent mechanism linking extreme mid latitude weather to AA and by linking differences in modeled responses to biases in the mean circulation state.
\end{abstract}


\section{Introduction}  

The Arctic is warming at an accelerated pace compared to the rest of the globe. This phenomenon, known as Arctic Amplification (AA), is a robust feature of anthropogenic climate change that has emerged from present-day observations and that is projected by climate models to further intensify by the end of the century \citep{IPCC_AR6_SYR_2023}. Both local and remote feedbacks contribute to the accelerated warming of the polar region, which mainly affects the lower levels of the atmosphere. These include surface albedo and temperature feedbacks---related to lapse-rate and radiative processes---as well as changes in the meridional transport of energy \citep{screen2010central,previdi2021arctic,caballero2025polar}.

A warmer Arctic leads to a reduced meridional temperature gradient near the surface, potentially modifying the characteristics of the midlatitude atmospheric circulation, with important impacts on densely populated regions. This mechanism, together with an apparent increase in cold-spell frequency in the midlatitudes during the early 21st century, has motivated numerous studies on the response of midlatitude atmospheric variability to Arctic warming, some of which have led to conflicting results over the past decades. In particular, \citet{francis2012evidence} proposed a mechanism by which a weaker meridional temperature gradient leads to a slower and wavier jet, more prone to atmospheric blocking onset. However, similar analyses by \citet{barnes2013revisiting} and \citet{screen2013exploring} found no significant increase in midlatitude waviness during the early 21st century.

Subsequent studies have emphasized how no robust signal of a midlatitude response to Arctic amplification has yet emerged from observational datasets. This might be due to the limited length of the available observational record, as highlighted by \citet{overland2016difficult} and \citet{blackport2020insignificant}. Moreover, more recent observational studies show that the previously reported increase in cold-spell occurrence disappears when the last 15 years of reanalysis data are included, indicating that the earlier trends were strongly influenced by internal atmospheric variability \citep{blackport2020insignificant,blackport2024models}.

To investigate circulation changes expected by the end of the century, several modeling studies have been conducted following the Polar Amplification Model Intercomparison Project (PAMIP) protocol \citep{smith2019polar}. The adoption of a common protocol has helped reduce the wide spread in simulated Arctic amplification across CMIP6 climate models \citep{cohen2020divergent}. These studies have identified a weak but statistically significant response to AA, consisting of a weakening and an equatorward shift of the midlatitude jet stream \citep{smith2022robust,screen2022net}. Further analyses of these experiments have identified an increase in Scandinavian blocking frequency \citep{riebold2023linkage,hay2025impact}, together with a small increase in blocking over northeast Asia \citep{zhuo2024response} and contrasting changes over the North Pacific \citep{streffing2021response}. Moreover, \citet{lewis2024response} report an increase in near-surface air temperature persistence in the North Atlantic.

The use of diverse methods and metrics to quantify circulation changes across the extensive literature on AA impacts on midlatitude circulation has resulted in apparently contrasting responses. Early studies focused on jet-stream waviness \citep{francis2012evidence,screen2013exploring,barnes2013revisiting}, while later work examined cold-spell trends \citep[e.g.][]{screen2015reduced,schneider2015physics}, Rossby wave phase speed \citep{riboldi2020linkage}, transient kinetic energy \citep{shaw2022midlatitude}, and several waviness metrics \citep{hassanzadeh2014responses,cattiaux2016sinuosity,huang2016local,francis2015evidence}. These different metrics capture distinct aspects of eddy variability and can therefore lead to diverging responses. Moreover, the relationship between the climatology of these metrics and persistent temperature extremes, or atmospheric blocking, is non-trivial \citep{nakamura2018atmospheric,barpanda2025local}. As a result, a physically consistent description of how AA, jet-stream waviness, mean circulation changes, and storm-track position and intensity combine to affect atmospheric blocking occurrence is still lacking.

Alongside more comprehensive climate models, idealized climate models have proven to be a useful tool to study circulation changes associated with AA. Such models have been used to investigate the competition between upper-tropospheric tropical heating and AA \citep{butler2010steady}, the response of the stratospheric circulation to AA \citep{wu2016response,mudhar2024exploring}, the effect of ice-constraining methods on the atmospheric circulation response \citep{shaw2022midlatitude,lewis2024assessing}, the role of lapse-rate feedbacks \citep{feldl2017atmospheric,yuval2020eddy}, changes to the polar radiative-advective equilibrium \citep{caballero2025polar}, barotropic eddy–mean flow feedbacks \citep{shaw2022midlatitude}, and the role of Rossby wave breaking location \citep{ronalds2019role}. Earlier literature also showed that atmospheric blocking can occur in aquaplanet simulations despite the absence of zonal asymmetries \citep{hu2008blocking}. In this context, \citet{geen2023explanation} showed that different waviness metrics applied to idealized AA experiments can lead to opposite trends. In turn, \citet{hassanzadeh2014responses} studied the response of atmospheric blocking to AA using a blocking index based on geopotential height, finding a decrease in blocking frequency. However, \citet{hassanzadeh2014responses} results may be affected by the experimental setup, which produces an unrealistic Arctic warming pattern extending far into the midlatitudes and influencing the circulation response through changes in static stability \citep{yuval2020eddy}.   

Most idealized modeling studies assume zonally symmetric background conditions \citep[e.g.][]{butler2010steady,hassanzadeh2014responses,feldl2017atmospheric,yuval2020eddy,shaw2022midlatitude,mudhar2024exploring}, despite evidence that circulation responses to AA depend on longitude, differing, for example, between the Atlantic and Pacific basins \citep[e.g.][]{streffing2021response,zhuo2024response,smith2019polar,ronalds2019role,screen2025causes}. In this context, localized midlatitude storm tracks have been shown to be an important factor in determining regions of blocking onset \citep{nakamura2018atmospheric,barpanda2025local} and should therefore be explicitly represented when assessing links between changes in waviness and changes in blocking frequency relevant for Earth’s atmospheric circulation.

In this study, we aim to clarify how changes in the mean circulation affect jet-stream waviness and blocking occurrence within a physically consistent framework, in which differences in transient kinetic energy, waviness, and Rossby wave characteristics play different but complementary roles in shaping the atmospheric blocking response to AA. This is done by using idealized simulations with two configurations: a zonally symmetric aquaplanet, and a zonally asymmetric one where we investigate the effect of a localized storm track on the blocking response. The idealized experiments are conducted using the Isca intermediate-complexity modeling system, described in more detail in Section~\ref{sec:experimental_setup}. To provide a physically consistent interpretation of changes in waviness and temperature extremes, we adopt the Traffic Jam theory framework introduced by \citet{nakamura2018atmospheric} to explain atmospheric blocking onset. In this framework, blocking occurs when local wave activity (LWA)---a measure of midlatitude jet waviness \citep{huang2016local}---exceeds a critical threshold referred to as the jet’s carrying capacity. This threshold can be computed from the quasi-geostrophic potential vorticity field, the stationary wave amplitude and the midlatitude wind field, thereby quantitatively linking changes in the mean atmospheric circulation to changes in blocking frequency \citep{yan2024traffic,barpanda2025local}. 

The paper is structured as follows. Section~\ref{sec:methods} presents the methods, including the experimental setup, and the definitions of the diagnostics used for blocking detection, carrying capacity and transient kinetic energy computation. Section~\ref{sec:results} presents the results, starting with the unperturbed atmospheric circulation of the symmetric and asymmetric configurations, followed by the zonal-mean and local zonal responses to AA, and concluding with an interpretation of the circulation changes in terms of carrying capacity. Section~\ref{sec:isca_conclusions} summarizes our main results. Lastly, in the Appendix, we report a comprehensive description of the Traffic Jam theory that can serve as reference for additional details on the performed analysis.

\section{Methods}
\label{sec:methods}
\subsection{Model and experiments}
\label{sec:experimental_setup}

The numerical experiments conducted in this study are based on the Isca intermediate-complexity model \citep{vallis2018Isca}. Isca is a flexible framework for idealized simulations of planetary atmospheres at varying levels of complexity and realism, built upon the modeling infrastructure of the Flexible Modeling System (FMS) developed at the Geophysical Fluid Dynamics Laboratory (GFDL).

Our model configuration closely follows \citet{kaspi2013role}, itself based on \citet{frierson2006gray}. The experimental configuration consists of an aquaplanet lower boundary with no land-sea contrast or orography, in which the surface is represented by a $1.5$ m mixed-layer ocean with a prescribed analytical, observation-based meridional heat flux adapted from \citet{merlis2013hadley}. The prescribed insolation is time invariant and set to its annual-mean value, with no diurnal or seasonal cycle. Radiation is represented with a two-stream gray scheme that distinguishes only between shortwave and longwave components, while moist convection is parameterized using a simplified Betts-Miller scheme \citep{frierson2007dynamics} configured as in \citet{o2008hydrological}. 

We conduct four experiments: two control baseline runs (hereafter BASE) and two experiments that include a simple AA-like forcing (hereafter AA). Each pair consists of a zonally symmetric run (SYM) and a zonally asymmetric simulation where a triangular heating is prescribed in the ocean mixed layer (ASYM), mimicking a western boundary current and breaking the zonal symmetry of the storm track (following \citealp{kaspi2013role}). The four simulations analyzed are therefore BASE SYM, AA SYM, BASE ASYM, and AA ASYM. Each experiment is integrated for 50 years at T85 resolution, with daily output saved at $1.5^\circ\times1.5^\circ$ horizontal resolution and 30 vertical sigma levels. Because our configuration is hemispherically symmetric, we analyze both hemispheres by mirroring Southern Hemisphere data onto positive latitudes and appending the resulting fields, obtaining a 100-year time series defined for the Northern Hemisphere only.

The localized heating imposed in the ASYM experiments is designed to mimic the effect of a western boundary current, such as that at the entrance of the Atlantic and Pacific storm tracks and it is imposed similarly to \citet{kaspi2013role}. This forcing enhances localized eddy generation and produces a midlatitudes storm track comparable to those observed in reanalysis. The heat flux is prescribed uniformly over a triangular region with vertices $(25^\circ\mathrm{N},75^\circ\mathrm{W})$, $(25^\circ\mathrm{N},100^\circ\mathrm{W})$, and $(50^\circ\mathrm{N},75^\circ\mathrm{W})$, with a magnitude of $850\ \mathrm{W\,m^{-2}}$. Unlike \citet{kaspi2013role}, who apply the triangular heating only in the Northern Hemisphere, we impose it symmetrically in both hemispheres. To prevent the introduction of a net global energy source, we compute the heat added along each latitude circle and subtract an equal amount distributed uniformly across all longitudes. For instance, at $25^\circ\mathrm{N}$ we integrate the heating between $100^\circ\mathrm{W}$ and $75^\circ\mathrm{W}$ and subtract the same total amount along the full latitude circle from $180^\circ\mathrm{W}$ to $180^\circ\mathrm{E}$.

To emulate the effects of polar-amplified warming, in the AA experiments we further prescribe a uniform surface heat flux of $30\ \mathrm{W\,m^{-2}}$ poleward of $72^\circ\mathrm{N}$ and $72^\circ\mathrm{S}$. Similarly to the asymmetric forcing, we integrate the imposed heat flux over the polar area and we remove the same amount of heat uniformly from the global surface, to avoid global warming. The imposed heat flux produces a polar warming slightly stronger than end-of-century projections under high-emission scenarios, which helps amplify the circulation response and facilitates the study of blocking changes. Sensitivity tests with alternative forcing magnitudes produce qualitatively similar results (not shown). The latitudinal boundary of this heating is chosen by matching the climatological Northern Hemisphere winter sea-ice extent (retrieved from NASA datasets): from the observed sea-ice area, we derive the latitude of a zonally symmetric circle with equal area. Although the surface energy balance produced by this method is not physically realistic \citep{lewis2024assessing,caballero2025polar}, the resulting warming pattern closely resembles that of more comprehensive models, giving us confidence that its influence on the mean circulation is sufficiently realistic for the purposes of this study.

\subsection{Local wave activity and carrying capacity}
\label{sec:LWA_carrying_capacity_def}

We use the Traffic Jam Theory of \citet{nakamura2018atmospheric} to interpret the blocking response to AA in our simulations. The Traffic Jam theory interprets blocking as an increase in LWA resulting from nonlinear interactions between waves and the mean flow.  In order to effectively describe this theoretical framework, several considerations are needed. In this Section, for the sake of conciseness we only provide the definitions of the key quantities adopted in the analysis. A comprehensive description of the computations and reasoning involved is provided in the Appendix. 

Intuitively, LWA measures the deviations of the quasi-geostrophic potential vorticity (QGPV) field from a wave-less reference state, which is computed from a snapshot of the QGPV field following the methodology introduced by \citet{nakamura2010finite}. A thorough description of the wave-less reference state $q_{ref}$  computation is provided in Appendix~\ref{sec:APPENDIX:local_wave_activity_definition}. LWA can therefore be expressed as:
\begin{equation}
\mathcal{A}(\lambda, \phi, z, t) \cos\phi 
= -a \int_{\phi}^{\phi(q_{ref})} \big[ q(\lambda, \hat \phi, z, t) - q_{\mathrm{ref}}( \phi, z, t)\big] 
\cos(\hat{\phi}) \, d\hat{\phi},
\label{eq:LWA_def}
\end{equation}
where $a$ is the radius of Earth, $\phi$ is latitude, $\lambda$ is longitude, $z$ is the vertical coordinate and $t$ is the temporal coordinate. In the following definitions, the LWA is density weighted and vertically averaged, therefore neglecting the height dependence. Note that once $q_{ref}$ is known, it is possible to invert the QGPV equation to define a reference state $(u_{ref},\theta_{ref})$, with $u_{ref}$ being the reference zonal wind and $\theta_{ref}$ being the reference potential temperature. 

The mathematical framework adopted by \citet{nakamura2018atmospheric} to interpret blocking onset and maintenance is analogous to the model of \citet{richards1956shock} for a traffic jam on a highway. In this analogy, LWA represents the density of cars, the LWA zonal flux represents the car density flux, and stationary wave crests act as traffic bottlenecks. Blocking emerges when an accumulation of LWA (car density) leads to a decrease in local wave activity flux (cars slow down due to the proximity to other vehicles), triggering a nonlinear mechanism that blocks the flow. 

Besides providing a physically coherent framework for blocking onset, a key advantage of the Traffic Jam theory is that it allows the identification of regions where blocking preferentially occurs from a set of basic fields, such as the wind field and potential temperature. In this way, blocking occurrence can be directly related to the mean atmospheric circulation, or, in our case, to changes in the mean atmospheric circulation induced by AA.

The fact that LWA is computed solely from the QGPV field means it inherits its conservation properties. One can therefore write a semi-empirical conservation equation that makes the analogy with Richards' traffic jam model quantitative; this budget equation is one-dimensional and reads:

\begin{equation}
\frac{\partial A}{\partial t}
= -\frac{\partial}{\partial x} \left[ \left( u_0 + c_{xg} - \alpha A \right) A \right]
+ S - \frac{A}{\tau},
\label{eq:tjt_eq}
\end{equation}
Where $u_0$ is the time-averaged reference wind, $c_{xg}$ is longitudinal component of the Rossby wave group velocity, $\alpha$ is a locally-defined negative correlation field between LWA and the eddy zonal winds, $S$ is a LWA source term and $\tau$ is a damping coefficient representing friction and baroclinic processes.  Note that the sum $u_0 + c_{xg}$ is hereafter called the Doppler-shifted Rossby wave group velocity. A detailed derivation of the budget equation and the computation of each term is provided in Appendix \ref{sec:APPENDIX:traffix_jam_theory}.

It is useful to spend a few more words on the physical meaning of the field $\alpha$, which will be important in the rest of the paper. $\alpha$ can be computed directly from the eddy zonal wind field and the LWA through the relation
\begin{equation}
    u_e = -\alpha A,
\label{eq:uealpha_mainpaper}
\end{equation}
where the term $u_e$, the eddy wind, is defined as the deviation of the zonal wind field from the reference zonal wind ($u_e = u-u_{ref}$). Eq.~\ref{eq:uealpha_mainpaper} describes an anti-correlation between local wave activity and eddy zonal winds, attributing to Rossby waves the same characteristic as vehicles: where traffic increases, drivers decelerate. The physical meaning of $\alpha$ can be better understood through a geometrical argument: where LWA increases, the flow acquires a meridional component, at the expense of the zonal component, hence locally and temporarily decreasing the zonal winds. In a baroclinic atmosphere, the torque associated with eddies is not entirely used in deflecting the jet stream, but is partially dissipated in diabatic processes, causing a reduction in $\alpha$. On the contrary, a more barotropic atmosphere is generally associated with a higher $\alpha$, reaching the  limit $\alpha = 1$ for an idealized barotropic flow. The climatology reported by \citet{barpanda2025local} in reanalysis supports this arguments, showing higher values of $\alpha$ in the tropical latitudes and lower values of $\alpha$ in the storm track region. 

Returning to the budget equation (Eq. \ref{eq:tjt_eq}), the term $F(x) =  A\left( u_0 + c_{xg} - \alpha A \right)$ represents a flux, namely the zonal LWA flux and it has a quadratic dependence on LWA. By computing the maximum of such a parabola we obtain the maximum flux, hereafter called the carrying capacity. It can be expressed as:

\begin{equation}
F_c(x) = \frac{(u_0 + c_{xg} - 2\alpha A_0(x))^2}{4\alpha},
\label{eq:Fc_local}
\end{equation}
where $A_0(x)$ is the stationary wave amplitude. The carrying capacity has an important physical meaning and it represents the maximum LWA flux beyond which blocking onset occurs and LWA begins to accumulate. It therefore discriminates between two regimes: (i) a regime in which $F(x) < F_c(x)$, where an increase in LWA translates into an increased flux; in the traffic jam analogy, this corresponds to a situation in which a higher car density leads to a higher car flux; (ii) a regime in which $F(x) > F_c(x)$, where an increase in LWA results in a lower flux and an upstream accumulation of LWA. In the traffic jam analogy, this second regime corresponds to the situation in which a higher car density aggravates the congestion, and it is analogous to blocking formation. It follows that the lower $F_c(x)$, the greater the probability of blocking occurring at a given LWA value. The carrying capacity can be computed directly from reanalysis or experimental data following the methodology of \citet{barpanda2025local}. A detailed description of the method used to compute the carrying capacity is provided in Appendix \ref{sec:APPENDIX:carrying_capacity_method}.

\subsection{Carrying capacity decomposition}
\label{sec:method_carrying_capacity_decomposition}
In this work we compare the carrying capacity of several experiments to understand how changes of the mean atmospheric circulation due to Arctic amplification lead to changes in blocking frequency.
To this aim, we decompose the total change of the carrying capacity in response to AA into contributions from $\alpha$, the negative correlation coefficient between LWA and eddy wind, the Doppler-shifted Rossby wave zonal group velocity  ($u_0+c_{xg}$) and the stationary LWA ($A_0$). This decomposition is expressed as:
\begin{equation}  
    \Delta F_c = \Delta F_{c,\alpha} + \Delta F_{c,u_0+c_{gx}} + \Delta F_{c,A_0} + residual \ .
    \label{eq:flux_decomposition}
\end{equation}
where each term on the right-hand side quantifies the specific contribution of a different physical quantity. For instance, the response of the carrying capacity $\Delta F_{c,\alpha}$ due to changes in $\alpha$ is computed as:
\begin{equation}
    \Delta F_{c,\alpha} = F_c(\alpha_f,(u_0+c_{gx})_b,A_{0,b}) - F_c(\alpha_b,(u_0+c_{gx})_b,A_{0,b}) \ ,
\end{equation}
where the subscripts \textit{b} and \textit{f}  denote 'baseline' and 'forced', respectively. The residual in Eq.~\ref{eq:flux_decomposition} is estimated by computing the total response of $F_c$ directly from the simulation output and then subtracting the other terms of the decomposition. For all simulations, the residual is negligible and, hence, not shown in the following analyses.

\subsection{Blocking detection and tracking}
\label{sec:block_detection_tracking}
We detect atmospheric blocking using a recent release of the \textit{blocktrack} algorithm, an open source tool designed to detect atmospheric blocking from gridded datasets \citep{filippucci2024impact}, that includes a new blocking index based on LWA. This new LWA-based index works similarly to widely adopted 500-hPa geopotential height anomaly-based indices, such as the one introduced by \citet{dole1983persistent}, but it is based on LWA. Specifically, for each day, we compute LWA anomalies $A'$ in the 40--80$^\circ$ latitudinal band. We then compute the standard deviation of these anomalies ($\sigma_{A'}$) identifying a single value for the whole time series and spatial domain. This approach is appropriate because our simulations do not feature a seasonal cycle or a time-varying external forcing. The same process is applied to each model configuration, defining a different threshold for each experimental setup. Finally, grid points where LWA anomalies $A'$ satisfy the criterion:
\begin{equation}
A' > 1.26 \sigma_{A'}
\label{eq:blockingcriteria}
\end{equation}
are identified as candidate blocked grid points. The threshold value corresponds to the 90th percentile of a Gaussian distribution. 

The LWA anomalies are then tracked in time using the blocktrack Lagrangian tracking. A detailed description of the tracking algorithm is provided in \citet{filippucci2026tracking}. We then filter the events selecting only those with area larger than $3\cdot 10^6\ \mathrm{km}^2$ and duration longer than 4 days. The persistence filter was chosen following established literature on the topic, as reviewed by \citet{woollings2018blocking}, while the area threshold was determined through sensitivity tests to match the climatological frequency of blocking identified by alternative indices (not shown). 

The composites shown in Section~\ref{sec:exp_mean_state} are computed in the reference system of the blocking event's center of mass to characterize the average properties of blocking events in the baseline simulation. More specifically, we analyze each blocking event separately, computing its center-of-mass trajectory and selecting an area of $70 ^{\circ}$ longitude $\times \ 40 ^{\circ}$ latitude centered over a grid point $10 ^{\circ}$ latitude south of its daily center of mass. The southward offset is introduced to better highlight the equatorward features of the blocking system that would otherwise be excluded from the composite. We repeat this procedure for all blocking events and we average the mass-weighted vertically averaged QGPV (hereafter just QGPV, unless otherwise stated) and the 850-hPa temperature anomaly over the inspected areas.
 
\subsection{Eddy kinetic energy}
\label{EKE_methods}
We use the 250-hPa eddy kinetic energy (EKE) to quantify the impact of the imposed asymmetric ocean heat flux relative to the zonally symmetric control. EKE is computed by applying a Fast Fourier Transform (FFT) to the zonal and meridional wind fields at 250 hPa and then applying a 2--6 days band-pass filter to isolate the high-frequency variability associated with transient eddies. Finally we  transform back the fields to the physical space. After applying the filter, EKE is computed as:
\begin{equation}
    EKE = \frac{u_{bp}^2+v_{bp}^2}{2}
\end{equation}
where $u_{bp}$ and $v_{bp}$ are the band-passed zonal and meridional winds, respectively. Compared to LWA, EKE typically measures the energy associated with short-lived eddies, such as those forming the storm track, whereas LWA does not distinguish between temporal scales of variability. Moreover, while EKE is computed at an upper-tropospheric level (250~hPa), LWA is a mass-weighted vertically averaged quantity.

\subsection{Statistical significance}

Statistical significance is assessed using a Student's t-test. The variance of the variables presented in the Results section is calculated through a bootstrap technique. The significance of the anomalies (both between ASYM and SYM simulations, and between AA and BASE simulations) is computed as:
\begin{equation}
    T_s = \frac{<x_1> - <x_2>}{\sqrt{\sigma_1^2 + \sigma_2^2}}
\end{equation}
where $T_s$ is the Student's $t$-value, $x_1$ and $x_2$ are the temporal means of the analyzed fields and $\sigma_1^2$ and $\sigma_2^2$ are their variances. The $T_s$ value is then compared with a reference value for the 95\% confidence interval. Since our experimental setup is computationally efficient, the number of simulated years was chosen to ensure that all anomalies shown below are significant at this confidence level.

\section{Results}
\label{sec:results}
\subsection{Baseline experiments: atmospheric circulation and blocking climatology}
\label{sec:exp_mean_state}

We begin our analysis by assessing whether our baseline experiments are able to represent the atmospheric circulation features we are interested in. In particular we examine the features of the storm track in both experiments, which we expect to be zonally symmetric in the BASE SYM experiment and asymmetric in the BASE ASYM experiment. Moreover, we highlight the differences between the EKE and LWA and their relation with the storm track.

\begin{figure}[htbp]
    \centering
    \includegraphics[width=1.0\textwidth]{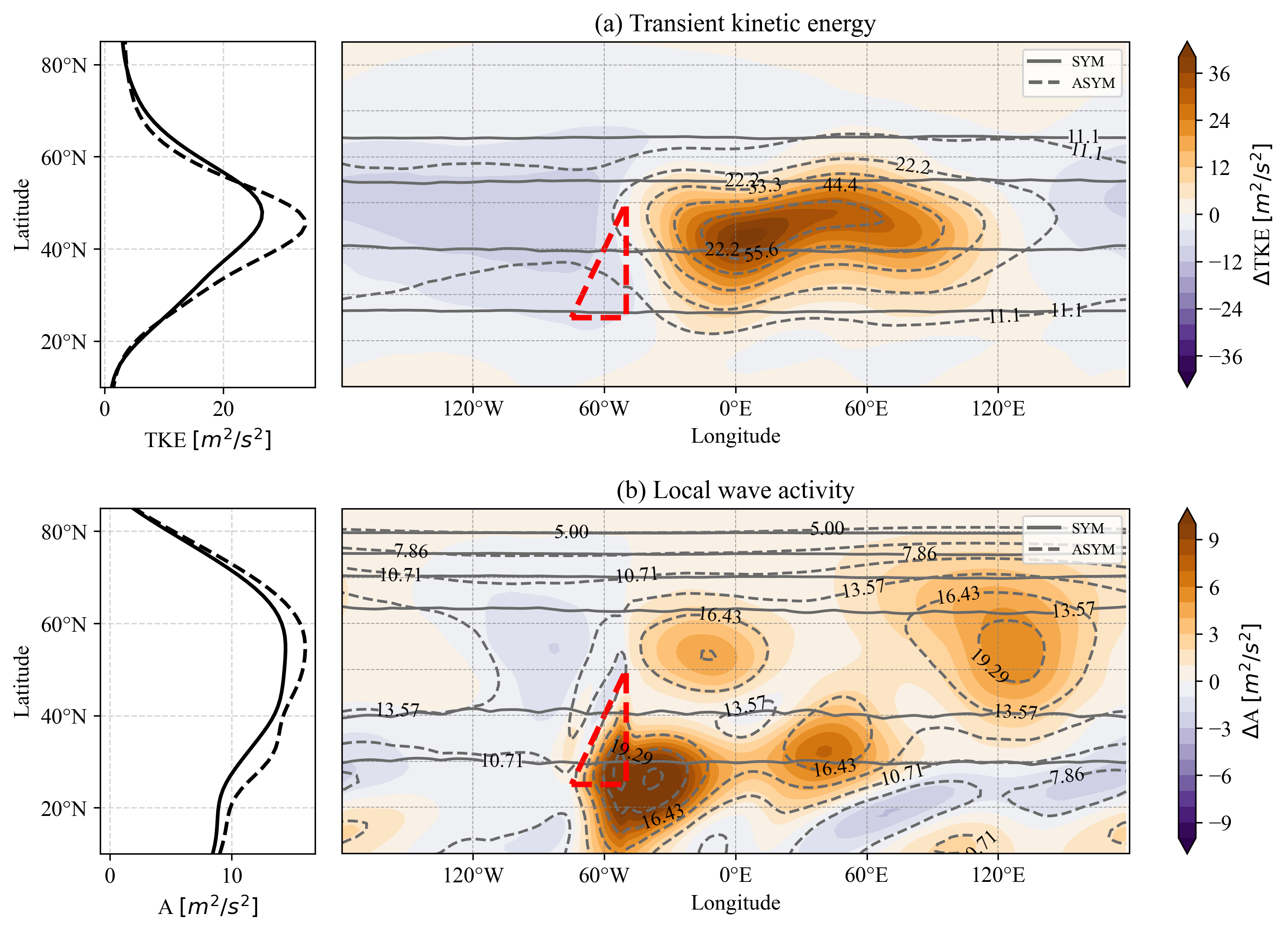}
    \caption{Climatology of EKE at 250hPa (panel a) and local wave activity (panel b).  In each panel dashed contours represent the BASE ASYM simulation, while solid contours represent the BASE SYM simulation. The shadings depict the differences between the two (BASE ASYM - BASE SYM). The red triangle identifies the region where the triangular ocean heat flux has been applied in the ASYM simulations. On the left of each panel, the zonal average of the BASE ASYM (dashed line) and BASE SYM (solid line) is reported.}
    \label{fig:EKE_lwa_climatology}
\end{figure}

We begin by comparing the EKE climatology in the BASE SYM and BASE ASYM simulations (Fig.~\ref{fig:EKE_lwa_climatology}a). The BASE SYM simulation exhibits a zonally symmetric EKE distribution with a maximum extending from 20°N to 60°N. In line with our expectations, in the BASE ASYM simulation, the imposed heat flux locally enhances transient eddies and their associated energy, producing a localized storm track.  

A comparison of the symmetric and asymmetric (Fig.~\ref{fig:EKE_lwa_climatology}a) setups reveals a spatially heterogeneous EKE response to the triangular heat flux: EKE increases downstream of the prescribed heating but decreases elsewhere. Nevertheless, the zonal mean shows an overall increase in EKE in the ASYM run. The localized enhancement is likely driven by increased baroclinicity, whereas the reduction elsewhere possibly results from the cooling applied along the same latitude band as the heating (see Section~\ref{sec:experimental_setup}). However, a similar EKE decrease was reported by \citet{kaspi2013role} even in the absence of such cooling, and was instead attributed to a stationary wave response to the localized heating. Notably, the asymmetric setup storm track shows good similarities to Earth's circulation over the Northern Hemisphere oceanic basins \citep{shaw2016storm} despite the idealized lower-boundary condition. 

We now shift our focus to LWA (Fig.~\ref{fig:EKE_lwa_climatology}b). We notice how the distribuiton of LWA and EKE is considerably different. This is in line with what anticipated in Section \ref{EKE_methods}:  EKE measures the energy associated with transient eddies, such as those forming the storm track, whereas LWA does not distinguish between temporal scales of variability. Moreover EKE refers to the upper troposphere, while LWA is a vertically integrated quantity.

In the BASE SYM simulation, LWA maximizes within a latitudinal band between 40°N and 60°N. In the BASE ASYM simulation, LWA exhibits multiple maxima downstream of the prescribed heating. The largest maximum, located at the south-east corner of the heat flux region (60°E, 25°N), likely arises from the sharp temperature gradient generated by the surface heat flux, which strongly influences low-level waviness. A similar interpretation applies to the maximum just downstream of the northern vertex of the triangular forcing (20°W, 50°N). The maximum at (50°E, 30°N) lies within the storm-track region identified in Fig.~\ref{fig:EKE_lwa_climatology}a and may reflect enhanced transient eddy activity in that area. Finally, the maximum near (120°E, 50°N) appears at the exit of the localized storm track and corresponds to the crest of a stationary wave induced by the zonal asymmetry. The correspondence was verified by plotting the zonal anomaly of the 500-hPa geopotential height with respect to the long-term mean (not shown), which shows a wave-number 1 stationary wave pattern, with its crest over (120°E, 50°N). This feature is analogous to the stationary wave crests observed over Northern Europe in reanalysis data (with different wave-number). In this case, however, the stationary wave arises solely from the imposed localized heating, rather than from land–sea contrasts or Earth's topography \citep{held2002northern}. This final LWA maximum is the most relevant for the purpose of this study, as it corresponds to the region of blocking onset, as discussed below.

\begin{figure}[htbp]
    \centering
    \includegraphics[width=1.1\textwidth]{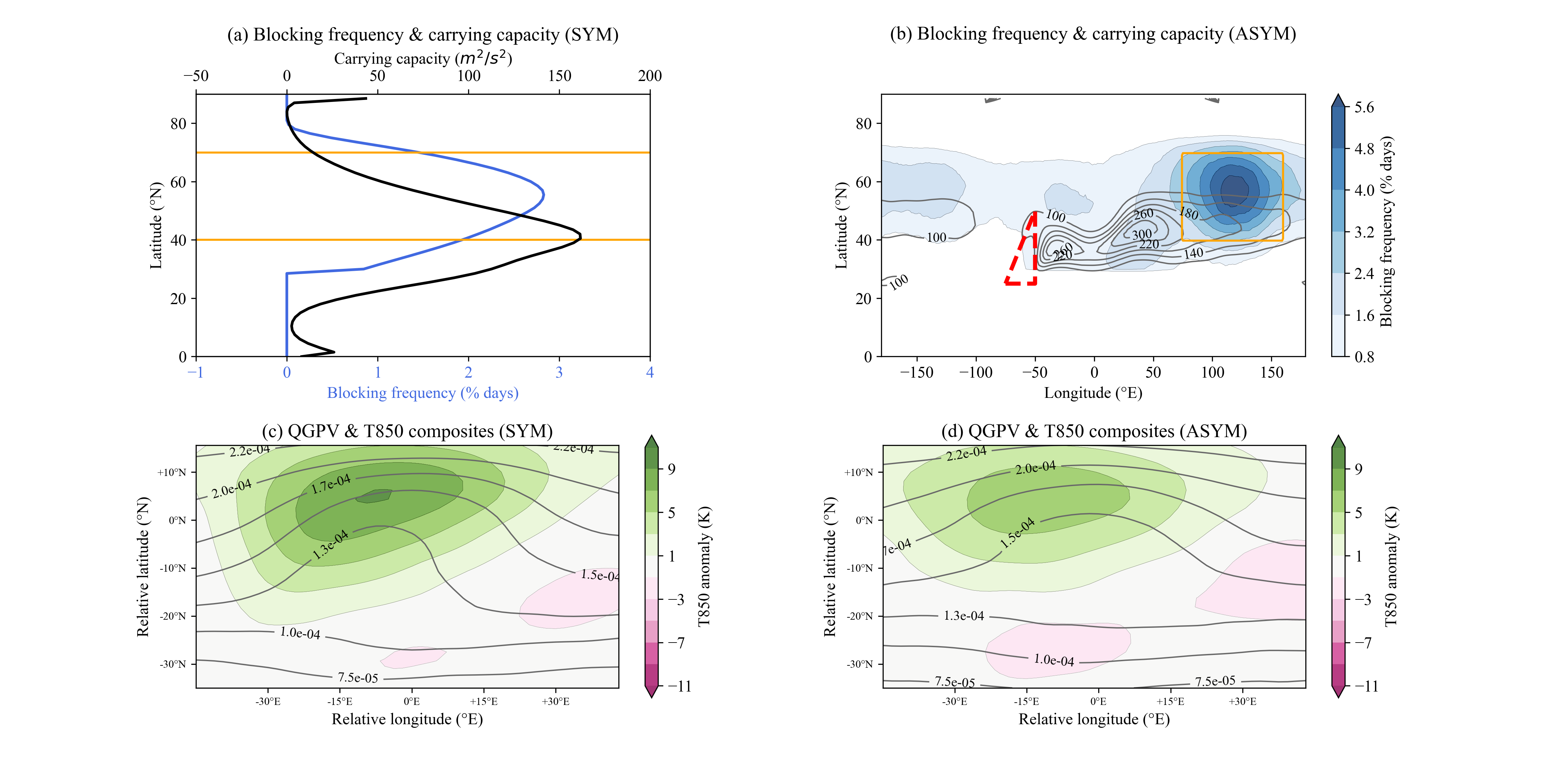}
    \caption{Panel (a) and (b) compare the atmospheric blocking frequency of the BASE ASYM and BASE SYM simulations with the corresponding carrying capacity. In panel (a) the zonal average of these two quantities for the BASE SYM experiment is plotted.  The black line refers to the zonally averaged carrying capacity and the blue line to the zonally averaged blocking frequency. In panel (b) the black contour represents the carrying capacity of the BASE ASYM experiment, while the blue shadings represent the blocking frequency. The red triangle identifies the region where the triangular ocean heat flux has been applied in the ASYM simulations. In both panel (a) and (b) the ocher boundaries delimit the area for which the composites of panel (c) and (d) respectively were computed. Panel (c) showcases the QGPV composite (black contours) and the 850 hPa temperature anomaly composite (colored shadings) for the BASE SYM simulation. Panel (d) showcases the same quantities for the BASE ASYM simulation.}
    \label{fig:blocking_climatology}
\end{figure}

Having investigated the LWA and EKE climatology, we evaluate the blocking climatology in the two baseline simulations (Fig.~\ref{fig:blocking_climatology}a,b) and compare it with the carrying capacity to evaluate the applicability of the Traffic Jam theory to our experimental setup. 

We start by analyzing Figure~\ref{fig:blocking_climatology}a, showing the zonal mean blocking frequency and carrying capacity for the BASE SYM simulation. The climatological maximum of blocking frequency occurs at 55°N, reaching approximately 2.5 blocked days per 100 days. As a reference, blocking frequency over Northern Europe in reanalysis data is typically assessed at 10 to 15\% of days, depending on the detection method used \citep{woollings2018blocking}, while in other regions, such as the Central Pacific, the blocking frequency is close to zero. Other analyses based on idealized aquaplanet simulations show blocking frequency similar to those found here \citep{narinesingh2020atmospheric,jimenez2022role}. The distribution of blocking is consistent with the distributions of LWA (Fig.~\ref{fig:blocking_climatology}a) and the carrying capacity. The blocking frequency is maximised close to 60°N, which corresponds to a region of relatively high LWA combined with reduced carrying capacity. South of this maximum, the blocking frequency decreases due to an increase in the carrying capacity, while, to the north, the blocking frequency decreases because the LWA is decreased.

Figure~\ref{fig:blocking_climatology}b illustrates the two-dimensional distribution of blocking frequency and carrying capacity for the BASE ASYM simulation. In this case, the blocking frequency maximum is located near (120°E, 55°N), reaching about 5 \% of blocked days over the simulated period. Notably, the latitude of this maximum remains the same as in the BASE SYM simulation. Furthermore, the blocking frequency increases downstream and along the northern flank of the carrying capacity distribution, reflecting the threshold mechanism underlying blocking onset in the Traffic Jam theory framework. The climatological blocking frequency maximum also coincides with the LWA maximum at (120°E, 50°N) identified in Fig.~\ref{fig:EKE_lwa_climatology}b, supporting our hypothesis that this region corresponds to the crest of a stationary wave induced by asymmetric heating---an area whose characteristics resemble the exit of the Northern Atlantic storm track in ERA5 reanalysis, where blocking preferentially occurs \citep{woollings2018blocking}. Moreover, blocking frequency is rather low between longitudes 180°W and 70°W, despite the reduced carrying capacity. This reduction likely results from the low climatological LWA in the region (see Fig.~\ref{fig:EKE_lwa_climatology}b), owing to the scarcity of LWA sources. Finally, we do not observe blocking in the south east corner of the triangular heat flux, despite the low carrying capacity and the high LWA observed in the region. This is likely explained by the fact that such a LWA maximum is not due to barotropic dynamics, but rather to the presence of a sharp temperature gradient at the surface. Moreover, south of 30°N, the quasi-geostrophic approximation at the core of the Traffic Jam theory loses validity. To conclude, the distribution of atmospheric blocking frequency is in line with what described by \citet{barpanda2025local} for ERA5 reanalysis: blocking onset occurs preferentially at the exit of the localized storm track, where carrying capacity values drop.

Summarizing, we find that in the BASE SYM simulation blocking occurs equivalently along all longitudes. In contrast, in the BASE ASYM setup the blocking frequency is not uniform, and a maximum emerges in correspondence of the stationary wave crest. This maximum can be explained by the lower carrying capacity in that region.

Fig.~\ref{fig:blocking_climatology}c,d shows composite plots of blocking events occurring in the BASE SYM and BASE ASYM simulations. To examine the characteristics of these events, we compute the 850-hPa temperature and QGPV, averaged over the lifecycle of each blocking event. As described in the Methods section (see Section~\ref{sec:block_detection_tracking}), the composites are presented in the reference frame of the blocking events' center of mass. 

Overall, the composites show characteristics similar to those of more comprehensive models, or reanalysis \citep{woollings2018blocking,sousa2018european}, even though the QGPV contours do not show the feature of a breaking Rossby wave, but rather those of a prominent and persistent ridge. Yet, the latter falls in the broad definition of a block \citep{woollings2018blocking}. Moreover, through a separate analysis we found that Rossby wave breaking (RWB) events are still occurring (not shown), both anticylonic and cyclonic, and the showcased QGPV contours likely reflect the fact that we are averaging both families of RWB events.

In both the BASE SYM and BASE ASYM cases, the prominent ridge identified by the tracking algorithm causes temperature anomalies as high as 9°K. Temperature anomalies are considerably larger in the BASE SYM simulation. Such large differences can be due to two different factors: (i) since the carrying capacity in Fig.~\ref{fig:blocking_climatology}b in the blocking frequency maximum region of the ASYM case is significantly lower than the zonally averaged SYM case, the LWA needed for blocking onset is smaller, which also translates in a less intense ridge. (ii) In the BASE ASYM simulation, atmospheric blocking occurs over a stationary wave crest, where the average 850-hPa temperature is higher. For the same LWA value, the temperature anomaly observed is therefore lower. Moreover, temperature anomalies are considerably larger in our idealized simulations than for ERA5 blocking events \citep[e.g.]{sousa2018european}. Such differences must be interpreted in light of the surface heat capacity, which in our idealized simulations is uniform and equal to a $1.5$-m deep water column. Such an idealized surface may have a thermal inertia considerably lower than that of the regions of Earth where atmospheric blocking normally occurs (e.g. Scandinavia, where the blocking center region is partially covered by the Nordic and Baltic seas), allowing temperature anomalies associated with the events to be significantly higher.

By inspecting the climatological LWA, EKE, carrying capacity and atmospheric blocking frequency we gain confidence that our model is able to reproduce the mechanism of blocking onset coherently with the Traffic Jam theory. In the next sections we will therefore assess the response of atmospheric blocking to AA, interpreting the changes in light of this theoretical framework.

\subsection{Forced local and zonally symmetric response}
\label{sec:forced_response}

\begin{figure}[htbp]
    \centering
    \includegraphics[width=0.8\textwidth]{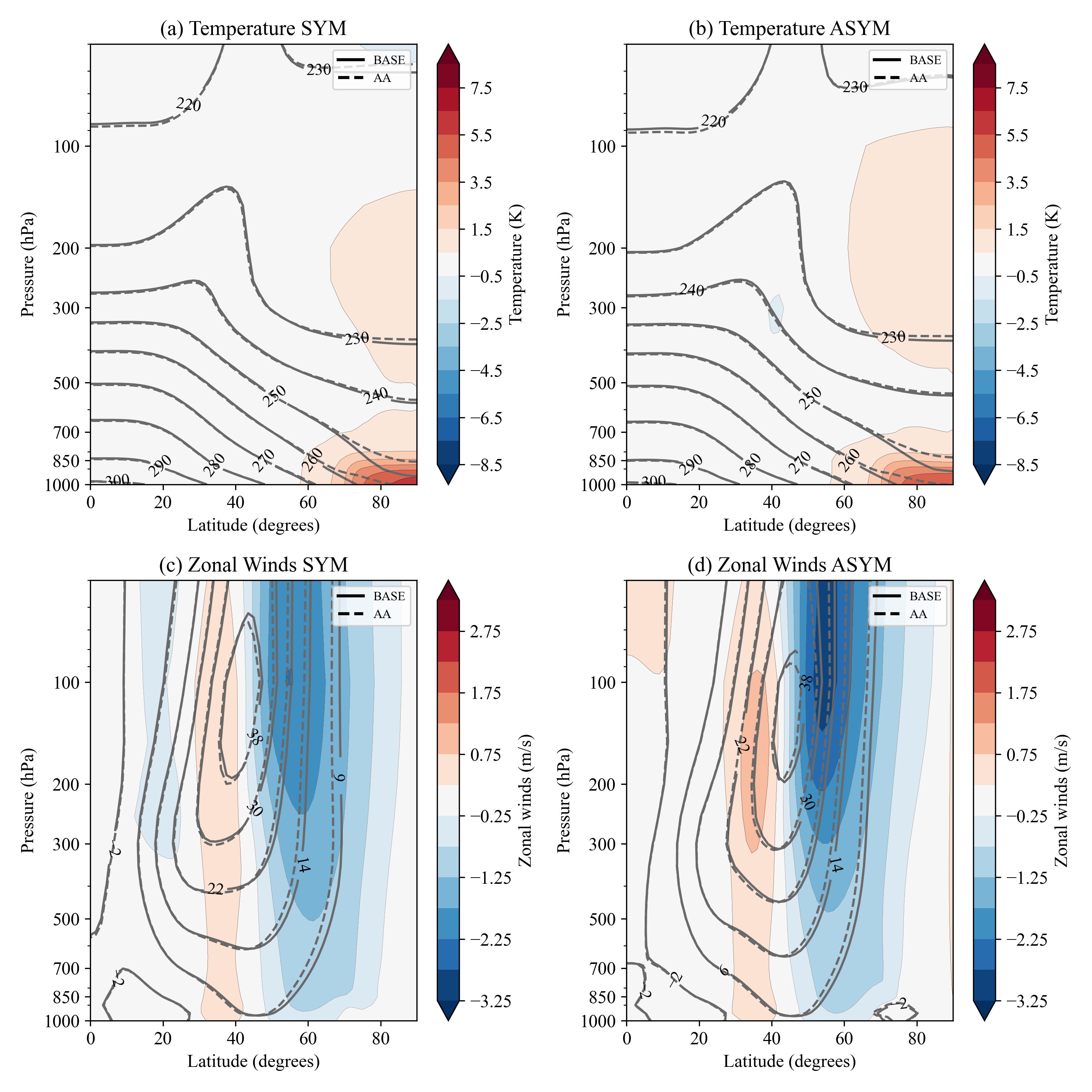}
    \caption{Panel a and b report the zonally averaged temperature response to AA, in the SYM and ASYM simulations respectively. Panel c and d depict the zonally averaged zonal winds response to AA, in the SYM and ASYM simulations respectively. In all plots, black solid (dashed) contours show the represented field as simulated by the BASE (AA) simulation. Moreover, shadings represent the difference between the AA runs and the BASE runs.}
    \label{fig:climatology_response}
\end{figure}

We begin by analyzing the zonally average zonal wind and temperature anomalies in the meridional plane to examine the response of the atmospheric circulation to AA (Fig.~\ref{fig:climatology_response}). The wind and temperature response show similar features across the symmetric and asymmetric experiments. As shown in Fig.~\ref{fig:climatology_response}a-b, for both the SYM and ASYM simulations, the temperature response exhibits a pronounced low-level warming beginning around 60°N and increasing toward the pole, reaching values of 5.5--6k over the Arctic. As anticipated in Sec.~\ref{sec:experimental_setup}, this warming is slightly higher than the AA projected by many current climate models under business-as-usual scenarios at the end of the 21st century, as well as the Arctic warming projected by PAMIP experiments --- around 4 degrees Kelvin surface warming in the zonal average \citep{smith2022robust,hay2024steady}. The warming extends vertically up to about 700 hPa, decreases aloft, and then intensifies again near the top of the troposphere, producing a double-maxima structure coherent with more comprehensive models \citep{smith2022robust}. Finally, the Arctic warming is slightly stronger in the SYM simulations.

Moving to panels c-d, we show the response of the zonally averaged zonal winds to AA. We find that in both the symmetric and asymmetric setups zonal winds are displaced equatorward, with the weakening on the poleward side generally larger than the strengthening on the equatorial side; in the ASYM simulations, both effects are further amplified. The jet displacement is consistent with what observed in PAMIP \citep{smith2022robust,screen2025causes}, even though the poleward weakening is more pronounced, likely because of the stronger Arctic heating. Earlier studies attributed changes in zonal-wind intensity to eddy-mediated feedbacks. In fact, a less baroclinic atmosphere generates fewer midlatitude eddies, which account for a substantial fraction of the meridional transport of zonal momentum in the global circulation \citep{andrews1987middle}. This reduction in eddy activity leads to an equatorward displacement of the jet.

It therefore emerges that our simple experimental setup is able to roughly reproduce the circulation response to AA that has robustly emerged from more comprehensive modeling experiments \citep{smith2019polar,smith2022robust}. This makes our following analysis relevant for the interpretation of observed and modeled responses of blocking to AA, despite the idealized setup.

\begin{figure}[htbp]
    \centering
    \includegraphics[width=1\textwidth]{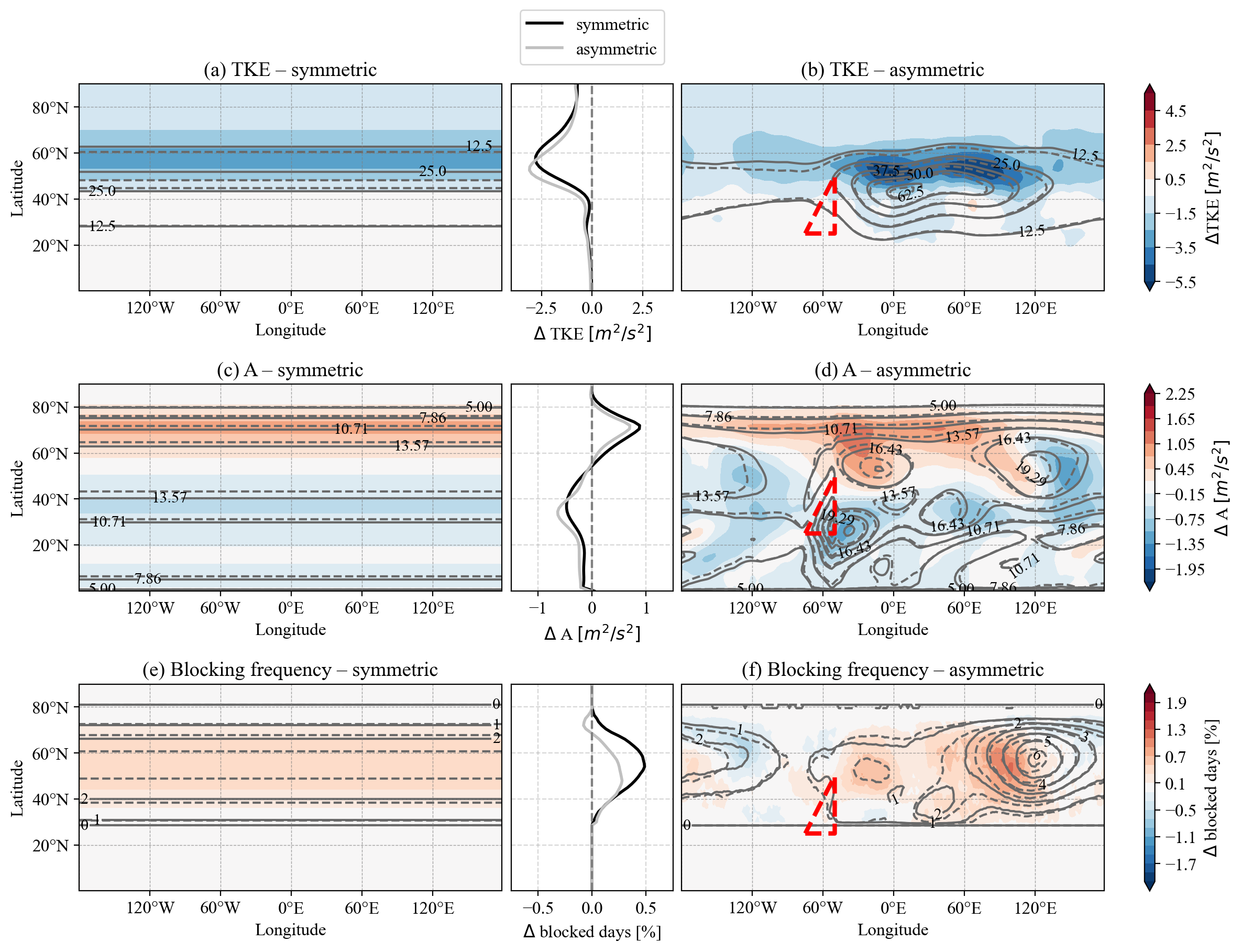}
    \caption{Panel a and b depict the response of EKE to AA, in the SYM and ASYM setups respectively. Panel c and d report the response of LWA, while panel c and f report the response of atmospheric blocking frequency. The shadings and the contours refer to the same experiments as in Fig.~\ref{fig:climatology_response}. In the central column, the zonally averaged differences between the two (BASE ASYM - BASE SYM) are reported, with the color scheme reported in the legend.}
    \label{fig:eddies_response}
\end{figure}

We now examine the response to AA of the more complex diagnostics analyzed in Section~\ref{sec:exp_mean_state}, needed to assess changes in jet stream waviness and heat extremes in the midlatitudes. Figure~\ref{fig:eddies_response} presents the behavior of EKE, LWA, and blocking frequency in both the SYM and ASYM experiments.

We find a reduction of EKE in both the symmetric and asymmetric setups (panels a and b). In the SYM simulations, EKE (first column; panel a) decreases by roughly 15--20\% along the northern edge of its climatological distribution, likely reflecting reduced baroclinicity at higher latitudes. In the ASYM simulations (last column; panel b), the reduction is more pronounced downstream of the imposed triangular heat flux anomaly, in the localized storm-track region, while it is less pronounced in areas where the climatological EKE is lower than the zonal average. Notably, the zonally averaged change (central column) is rather similar in the SYM and ASYM configurations.

Moving to the LWA response (panels c and d), we observe an equatorward displacement of its climatological distribution. Such a displacement mirrors changes in the zonal wind field, consistent with the negative correlation between LWA and eddy zonal winds (see Section~\ref{sec:LWA_carrying_capacity_def} and Appendix \ref{sec:APPENDIX:traffix_jam_theory}). An alternative explanation for this shift is provided by the interpretation of the jet stream as a waveguide that advects synoptic eddies \citep{platzman1968rossby,swanson1997dynamics,schwierz2004forced,martius2010tropopause}; according to this interpretation, a displacement of the jet stream is inevitably followed by a displacement of the eddy distribution, which in turn influences the climatological LWA. The LWA response to AA is broadly similar for the SYM (first column; panel c) and ASYM (last column; panel d) configurations, with peaks of 10--15\% relative change. Again, the zonally averaged response is similar between the two.

Lastly, the response of blocking frequency to AA differs between the SYM and ASYM configurations. In the SYM case (first column; panel e), we find a uniform increase in atmospheric blocking frequency in the latitudinal range $(45°N$--$65°N)$. Changes are rather large, peaking at $0.5\ \%\ days$, which corresponds to a $25\%$ relative change compared to the BASE configuration. Conversely, in the ASYM case (last column; panel f) we observe both regions of increased and decreased atmospheric blocking frequency, resulting in a displacement of the BASE configuration atmospheric blocking frequency maximum. Blocking frequency increases are larger and spread over larger areas than frequency decreases, resulting in a zonally averaged increase. Again, changes are rather large, peaking at $20$--$30\%$ relative change. Comparing the zonal response in the SYM and ASYM setups (central column), we find that the zonally averaged blocking increase is larger in the SYM setup than in the ASYM setup.

To summarize, while changes in LWA and EKE show comparable features in the SYM and ASYM simulations, the response of atmospheric blocking to AA exhibits important differences: while the SYM case exhibits a uniform increase, the ASYM response is better described as an upstream displacement of the blocking maximum and a modest increase in the zonal average.

\subsection{Interpretation of atmospheric blocking frequency changes}
\label{sec:interpretation}

To understand the mechanism behind the changes in blocking frequency described above, we analyze the physical quantities introduced in Section~\ref{sec:LWA_carrying_capacity_def}  within the context of the Traffic Jam theory. 

\begin{figure}[htbp]
    \centering
    \includegraphics[width=1\textwidth]{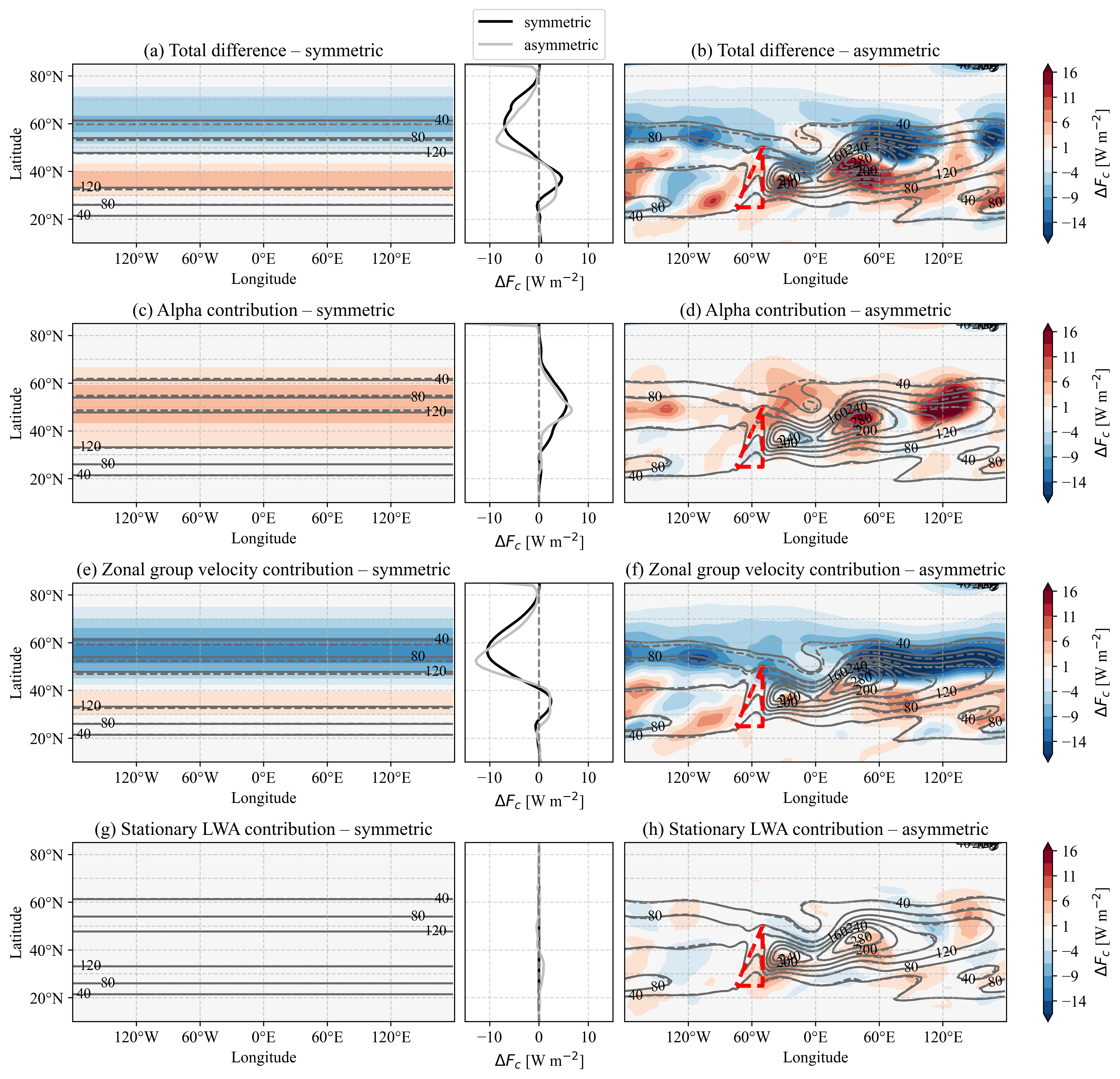}
    \caption{The Figure depicts the response of the carrying capacity to an AA-like forcing, in the SYM and ASYM simulations. Panels a-c-e-g refer to the SYM simulation, while panels b-d-f-h refer to the ASYM simulation. The response of carrying capacity is decomposed in various components: panel a and b show the total difference, panel c and d show the contribution of the $\alpha$ parameter, panel e and f show the contribution of the doppler-shifted Rossby waves group velocity and panel g and h show the contribution of stationary LWA. For further explanations about the computation of these contributions see Section~\ref{sec:APPENDIX:carrying_capacity_method} . The shadings and the contours refer to the same experiments as in Fig.~\ref{fig:eddies_response}}
    \label{fig:carrying_capacity_response}
\end{figure}

Figure~\ref{fig:carrying_capacity_response}a,b shows the response of the flow carrying capacity to AA in the SYM and ASYM model configurations. In both configurations the predominant signal is a strong carrying capacity decrease at higher latitudes and an increase at lower latitudes, with the response changing sign around $50°N$. While in the SYM case the depicted response is zonally symmetric, in the ASYM configuration we observe zonal asymmetries. In particular, changes in the carrying capacity are larger downstream of the triangular heat flux rather than upstream. We also observe a slight increase in carrying capacity at $120°E$, in correspondence of the stationary Rossby wave crest mentioned in Section~\ref{sec:exp_mean_state}. The zonally averaged response is similar between the SYM and ASYM configurations.

Such changes of carrying capacity are broadly compatible with the atmospheric blocking frequency changes. In particular, the reduction of carrying capacity poleward of $50°N$ is consistent with an increase of blocking frequency: a lower carrying capacity implies a decrease of the LWA zonal flux threshold after which blocking onset occurs, hence an higher frequency of atmospheric blocking.

Yet, the simple observation of the carrying capacity response to AA is not sufficient to explain the upstream displacement of the blocking frequency maximum in the ASYM simulations. The small increase at $120°E$ in panel b is not co-located with the blocking frequency decrease found in Fig.~\ref{fig:eddies_response}f, which is rather observed further east. In order to provide an explanation for the displacement it is necessary to consider the spatial structure of the climatological carrying capacity in the ASYM experiments: blocking onset preferentially occurs downstream and poleward of the carrying capacity maximum. Therefore, when the maximum weakens---even in the hypothetical case of a zonally symmetric carrying capacity response---the region where blocking can develop is displaced upstream, shifting the blocking frequency maximum accordingly. In this way, qualitatively similar circulation changes can produce distinct blocking responses depending on the underlying symmetries of the model.

Shifting our focus to Figure~\ref{fig:carrying_capacity_response} panels c,d,e,f,g,h we evaluate which physical variable involved in the carrying capacity definition is responsible for the observed response. Overall, we find that in both the SYM and ASYM configuration the $\alpha$ contribution (panels c, d) opposes the Doppler-shifted Rossby wave group velocity contribution (panels e, f), with the latter being larger than the former and shaping the overall sign of the response to AA; In both cases the zonally averaged responses (central column) is similar. Panel d shows how changes in $\alpha$ are the main reason for the zonal asymmetries in the ASYM configuration carrying capacity response to AA. Conversely, the contribution of the Doppler-shifted Rossby wave group velocity in the ASYM setup (panel f) is rather zonally symmetric, especially downstream of the triangular heat flux. Lastly, the contribution of $A_0$ is null in the SYM case and rather small in the ASYM case.

\begin{figure}[htbp]
    \centering
    \includegraphics[width=1\textwidth]{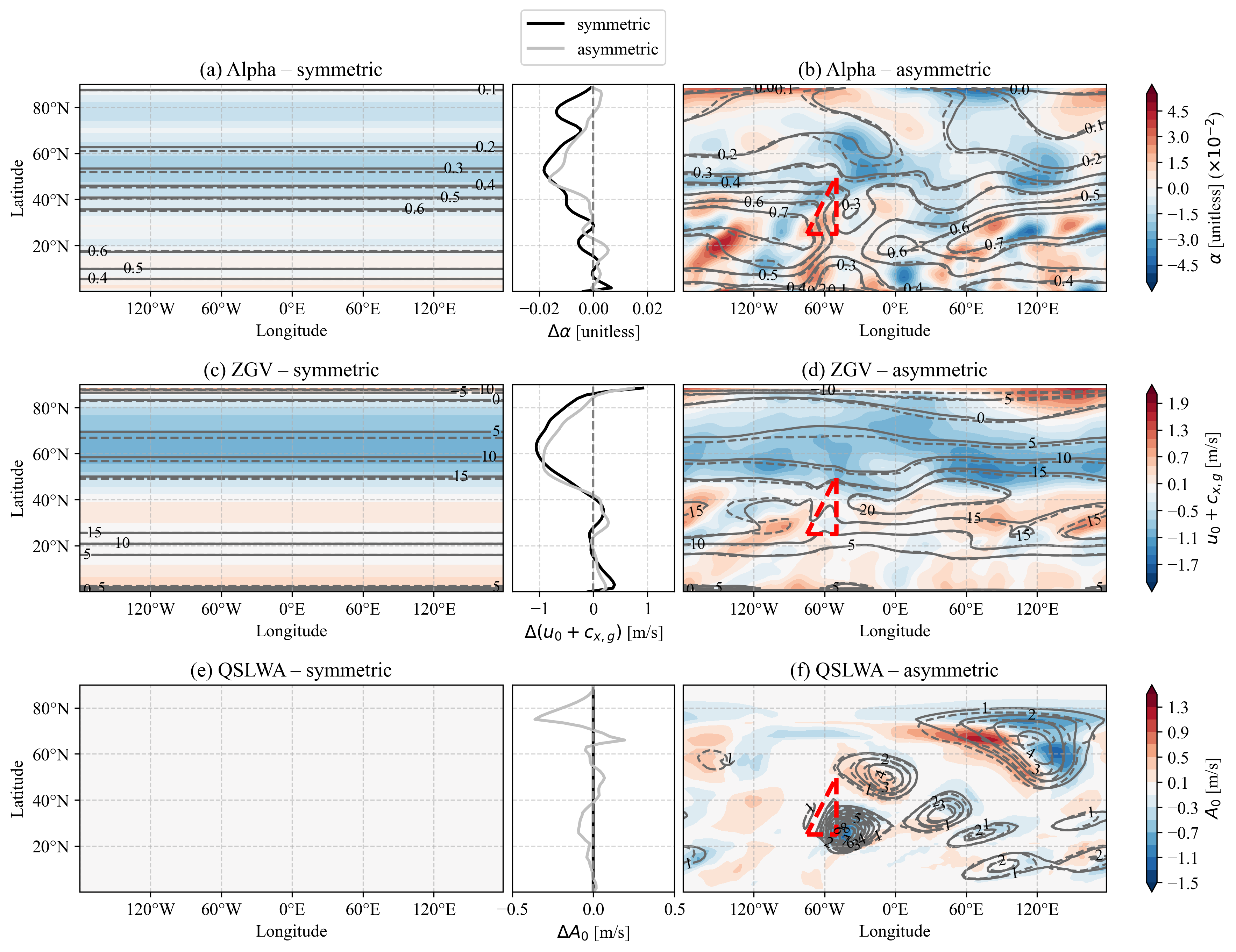}
    \caption{The Figure depicts the response of $\alpha$ (panels a,d), of the zonal Rossby wave group velocity (panels b,e) and of the quasi stationary local wave activity (panels c,f) to an AA-like forcing, in the SYM and ASYM simulations. Panels a-c-e-g refer to the SYM simulation, while panels b-d-f-h refer to the ASYM simulation. The shadings and the contours refer to the same experiments as in Fig.~\ref{fig:eddies_response}}
    \label{fig:alpha_zgv_qglwa_response}
\end{figure}

In order to better understand how these carrying capacity changes are connected to the mean atmospheric circulation response, it is useful to examine the response to AA of $\alpha$, $u_0 + c_{xg}$ and $A_0$. Fig.~\ref{fig:alpha_zgv_qglwa_response} shows the sensitivity of these key dynamical parameters across both experimental setups. 

Starting with the $\alpha$ parameter (panels a,b), we find that AA induces a decrease in both the SYM and ASYM simulations, with a slightly more pronounced zonally averaged decrease in SYM. According to the theoretical framework discussed in Appendix~\ref{sec:APPENDIX:traffix_jam_theory}, variations in $\alpha$ are expected to arise from baroclinicity changes \citep{nakamura2024large,barpanda2025local}, whereby a more baroclinic atmosphere features smaller $\alpha$ values, while more barotropic conditions lead to an increase. However, our results deviate from these expectations: despite the clear signals of reduced baroclinicity, manifested as diminished transient kinetic energy and a weakened meridional temperature gradient (Figures \ref{fig:eddies_response}, \ref{fig:climatology_response}), the parameter $\alpha$ decreases.

To identify the responsible mechanisms, we use the definition of $\alpha$ given in Section~\ref{sec:LWA_carrying_capacity_def} (or see Appendix \ref{sec:APPENDIX:carrying_capacity_method}, Eq.~\ref{eq:alpha_regression}) to decompose its change into contributions from changes in LWA variance and changes in covariance between LWA and zonal winds. In the Supplementary Material, we demonstrate that changes in the meridional wind shear may act to narrow the jet stream, concentrating the local wave activity variability in a narrower region and facilitating Rossby wave breaking on its northern and southern flanks, similarly to what reported by \citet{ronalds2019role}. These circulation shifts modify both the variance and covariance terms, ultimately resulting in a significant reduction of $\alpha$. Changes in the meridional wind shear are therefore expected to be of primary importance to assess the response of $\alpha$ to AA, while baroclinicity changes exert only a secondary effect. While a more detailed analysis of this mechanism is left for future work, a more comprehensive discussion and a series of supporting plots are provided in the Supplementary Material.

Moving to Fig.~\ref{fig:alpha_zgv_qglwa_response}c and d, we observe a weakening of the Doppler-shifted zonal Rossby wave group velocity following AA and a small strengthening southward of 40°N. This pattern of change depends both on changes of $u_0$, the temporal average of reference zonal winds (see Section \ref{sec:LWA_carrying_capacity_def}), which is a zonally symmetric quantity by definition, and changes of $c_{x,g}$, the zonal Rossby wave group velocity, responsible for the zonal asymmetries in the ASYM setup. An inspection of the temporal average of reference zonal winds (not shown) reveals that they are the main responsible for the Doppler-shifted zonal Rossby wave group velocity change, coherently with the fact that the ASYM setup shows an almost zonally symmetric response. Time-averaged reference zonal winds are in turn closely related to climatological zonal winds, which indeed we found to decrease at higher latitudes and slightly increase at lower latitudes with AA, as seen in Fig.~\ref{fig:climatology_response}. Lastly, the zonally averaged response is rather similar in the SYM and ASYM configurations.

To conclude, panels e and f show the stationary LWA response. In the SYM case, we do not display stationary wave anomalies, as these simulations do not feature zonal asymmetries. In the ASYM case, stationary LWA maxima move upstream following AA. This shift again follows the modification of the zonal winds: according to the linear Rossby wave dispersion relation, the wavenumber of stationary Rossby waves increases for stronger zonal winds. Conversely, slower winds imply shorter stationary Rossby wave wavelength. This translates into an upstream displacement of the stationary Rossby wave crest with respect to the imposed heating, as clearly evident in panel f.

To conclude, we find that the carrying capacity response is primarily driven by changes in the Doppler-shifted Rossby wave group velocity, which in turn depends on the climatological zonal winds response. Changes of $\alpha$ act as a negative feedback, reducing the carrying capacity decrease at higher latitudes and limiting the atmospheric blocking response. Lastly, changes in $A_0$ are small and play a secondary role.
In simple words, in this Section we highlight the physical mechanism by which a decrease and equatorward displacement of mean zonal winds lead to modifications of atmospheric blocking frequency, consisting of a zonally averaged increase at mid-to-high latitudes and an upstream displacement of the frequency maximum.

\section{Conclusions}
\label{sec:isca_conclusions}

In this study, we investigate the response of atmospheric blocking to Arctic Amplification (AA) within the framework of the Traffic Jam Theory. The use of idealized aquaplanet simulations in both zonally symmetric and asymmetric configurations, allows us to isolate the role of zonal asymmetries and baseline circulation features in shaping the blocking response to AA.

We first show that the experimental setup realistically reproduces key ingredients of midlatitude dynamics, namely storm tracks, LWA, and atmospheric blocking. Consistent with previous studies, blocking-like circulations emerge even in the absence of zonal asymmetries in the SYM simulations \citep{hu2008blocking,jimenez2022role}. The introduction of a localized storm track in the ASYM case leads to a spatial distribution of blocking that resembles observations in the Atlantic basin, validating the model's ability to capture essential processes governing blocking onset.

An important finding of this study is that despite similar large-scale circulation responses to AA in the SYM and ASYM configurations, manifest in comparable shifts in zonally averaged LWA and EKE in both configurations, the resulting blocking response differs markedly. In the SYM configuration, AA leads to a relatively uniform increase in blocking frequency at high latitudes. In contrast, the ASYM configuration is characterized by a longitudinal reorganization, where the dominant signal is an upstream shift of the blocking maximum.

We interpret these divergent changes in terms of changes in the carrying capacity. Under AA, the carrying capacity decreases at high latitudes and increases at mid-to-low latitudes, primarily as a consequence of changes in Doppler-shifted Rossby wave group velocity. While these carrying-capacity changes have similar characteristics in both configurations, their impact on blocking is dictated by the baseline climatological carrying capacity. In particular, in the ASYM case, the pre-existing spatial structure of the carrying capacity implies that a high-latitude decrease shifts the location where the conditions for blocking onset are met. This results in an upstream displacement, rather than a simple increase in frequency. 

A key contribution of this study is the application of the Traffic Jam theory of blocking onset to the response of midlatitude circulation to AA. This framework is particularly useful for explaining how changes in the mean atmospheric circulation, local wave activity, stationary waves, and transient kinetic energy combine to shape the blocking response in a physically consistent manner. In particular, we find that the decrease in zonal winds is the primary driver of the high-latitude increase in atmospheric blocking, with the stationary wave playing a secondary role. More in general, our results support the hypothesis that Arctic amplification can alter the midlatitude atmospheric circulation, weakening the zonal winds and increasing the probability of occurrence of temperature extremes \citep{francis2012evidence}. Furthermore, we identify a novel negative feedback associated with a reduction in $\alpha$, the correlation parameter between zonal winds and LWA. This reduction effectively buffers the decrease in carrying capacity, limiting the potential increase in blocking under AA. This mechanistic perspective, which accounts for several competing processes influencing blocking frequency under AA, provides a useful and unifying framework to contextualize and interpret previous studies that focus on different metrics related to Rossby wave characteristics, jet waviness, and midlatitude temperature extremes.


Furthemore, our results offer a physical explanation for the persistent inter-model spread in blocking projections. Specifically, our finding that the presence of a localized storm track fundamentally alters the response suggests that model biases in the mean-state carrying capacity can lead to considerably different blocking trends, even when the external forced changes in zonal winds, EKE and LWA are similar. This perspective provides further support to earlier arguments emphasizing the need to correct mean-state biases to improve the representation of blocking in climate models \citep{scaife2010atmospheric,filippucci2024impact}. Our study provides a physical explanation for this sensitivity, demonstrating that the unperturbed position and structure of the carrying-capacity play a central role in determining how blocking responds to external forcings such as AA.

While our study focuses on how mean-state changes impact blocking occurrence, previous research has investigated the role of Rossby wave breaking in the circulation response to AA \citep{ronalds2019role}. The relationship between mean atmospheric circulation and blocking is in fact twofold: the strength and position of the jet stream influence the location, frequency, and persistence of blocking events, while an increase in Rossby wave breaking---often associated with blocking occurrence \citep[e.g. ]{tibaldi1990operational,scherrer2006two,davini2012bidimensional}---can decelerate the high-latitude flow. This creates a “chicken-and-egg” problem, complicating the distinction between whether mean circulation changes drive blocking frequency modifications or vice versa. Sensitivity tests we have conducted by excluding blocked grid points in carrying capacity circulation changes show that our results remain qualitatively unchanged (not shown), supporting the conclusion that it is primarily the mean circulation that affects blocking frequency. Nevertheless, blocking changes can act as positive or negative feedback mechanisms on the mean wind response.



In conclusion, there are several opportunities for future research motivated by our study. For example, extending the carrying capacity framework to the ensemble of comprehensive climate model simulations contributing to PAMIP \citep{smith2019polar} would facilitate a more robust interpretation of model responses to Arctic warming in future projections. A similar analysis could be applied to the CMIP6 \citep{eyring2016overview} archive to understand the broader effect of greenhouse gas forcing on blocking through our novel carrying capacity decomposition. However, these tasks pose significant challenges, primarily due to the difficulty of computing QGPV from datasets with relatively low vertical resolution, such as those available in the PAMIP and CMIP6 ensembles.

In addition, the sensitivity of the parameter $\alpha$ to external forcings deserves further investigation. In our simulations, $\alpha$ acts as a negative feedback that buffers the response of the carrying capacity under Arctic amplification; therefore, accurately quantifying its sensitivity is of primary importance for predicting future blocking frequency changes. In the present study we suggest that its response is related to changes in the meridional wind shear. Yet, the response of $\alpha$ to Arctic amplification remains underexplored, particularly when compared with the extensive literature on changes in mean zonal winds or stationary waves. Addressing this gap is essential to improve our understanding of the nonlinear response of temperature extremes to Arctic amplification and, more broadly, to anthropogenic global warming.

\appendix

\section{Local wave activity definition}
\label{sec:APPENDIX:local_wave_activity_definition}

A key quantity analyzed in this study is the local wave activity ($A$ or LWA), defined as in \citet{huang2016local}. Intuitively, $A$ measures the amplitude of the Rossby wave as the meridional displacement of Quasi-Geostrophic Potential Vorticity (QGPV) from a zonally symmetric wave-free state.

\begin{figure}[htbp]
    \centering
    \includegraphics[width=0.7\textwidth]{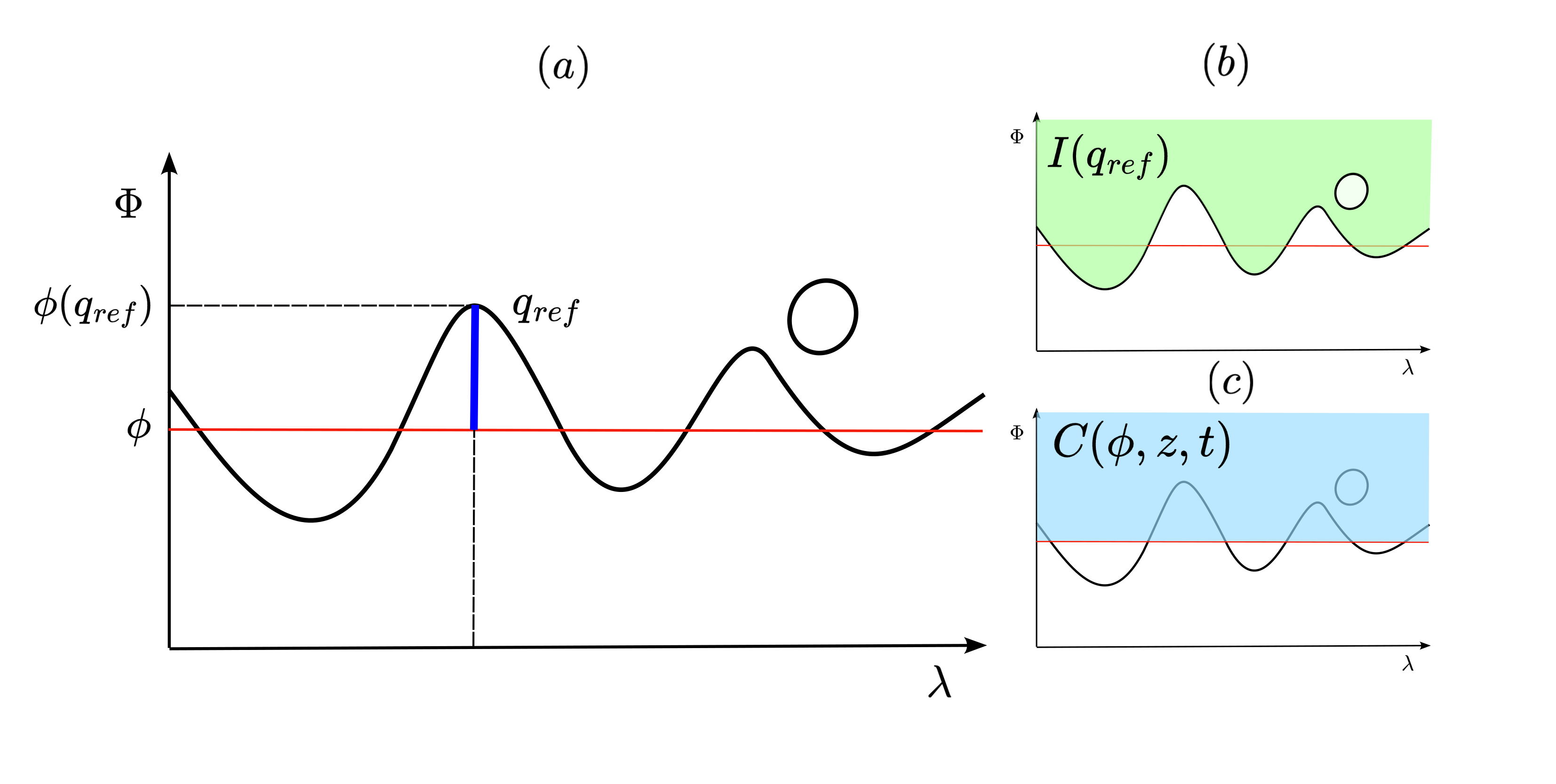}
    \caption{Schematic diagram showing (on a longitude-latitude plane) the variables and integration domains used in the LWA calculation. In panel a, the horizontal red line represent the i-th investigated latitude, $\phi_i$. The solid black contour represents the contour associated to the reference QGPV, $q_{ref,i}$. The solid blue meridional segment represents the domain over which the QGPV anomaly is integrated in Eq.~\ref{eq:LWA_def_APP}. In panel b, the green area represents the integral $I(q_{ref,i})$, delimited by the wavy reference contour. In panel c, the blue area represents the integral $C(\phi_i,z,t)$, defined as the area poleward of the fixed latitude $\phi_i$. By construction, the areas in (b) and (c) are equal.}
    \label{schematic:contoursQref}
\end{figure}

To compute this, we first define a zonally symmetric reference state $q_{ref}(\phi,z,t)$, where $\phi$ is the latitude, $z$ is the pseudo-height coordinate and $t$ is the temporal coordinate. The goal is to find a unique PV value, $q_{ref}$, for every latitude $\phi_i$ such that the meridional integral of PV poleward of the $q_{ref}$ contour, $I(q_{ref})$, is exactly equal to the meridional integral of the actual PV field poleward of the latitude $\phi$, $C(\phi,z,t)$. A graphical representation of variables and of the integrals $C(\phi,z,t)$ and $I(q_{ref})$ is shown in Fig.~\ref{schematic:contoursQref}. In equations this can be expressed as follows: considering a latitude $\phi$, we evaluate the integral
\begin{equation}
    C(\phi,z,t) \equiv \int_{-180°}^{+180°} \int_{\phi}^{90°} q(\lambda,\hat\phi,z,t) \cos(\hat{\phi}) \ d\hat\phi d\lambda,
    \label{eq:equivalent_latitude}
\end{equation}
where $\lambda$ is the longitudinal coordinate,  $\hat\phi$ is the latitudinal coordinate used for the integration and $q(\lambda,\hat\phi,z,t)$ is the QGPV field. We then determine the reference PV values associated to $\phi$,  $q_{ref}$ (or $q_{ref}(\phi,z,t)$), by evaluating the integral $I(q_{ref})$, 
\begin{equation}
   I(q_{ref}) \equiv   \int_{-180°}^{+180°} \int_{\phi(q_{ref})}^{90°} q(\lambda,\hat\phi,z,t)\cos(\hat{\phi})\  d\hat\phi d\lambda,
\label{eq:integralI}
\end{equation}
where $\phi(q_{ref})$ is the latitude of the contour of $q_{ref}$ at a longitude $\lambda$---and imposing the equivalence

\begin{equation}
    C(\phi,z,t) = I(q_{ref}). 
    \label{eq:C=I}
\end{equation}

Note that if at a given longitude, more than one contour is found, the area in the enclosed contour is removed from the integral defined in Eq.~\ref{eq:integralI}. This equivalence leads to a one-to-one map $q_{ref}(\phi,z,t)$. Finally, once $q_{ref}$ is known, it is possible to invert the QGPV equation to define a reference state $(u_{ref},\theta_{ref})$.

The procedure for calculating the reference state may be better understood by describing the algorithm used to compute it on a gridded dataset. First, for each element of a latitude array, $\phi_i$, the integral in Eq.~\ref{eq:equivalent_latitude}, $C(\phi_i, z, t)$, is evaluated. To find the corresponding $q_{ref,i}$, a QGPV threshold is initialized starting from the minimum QGPV value found at the analyzed time $t$, and the integral is computed over all grid cells with QGPV values less than or equal to the current threshold. This procedure is repeated while progressively increasing the threshold until the evaluated integral equals or exceeds $C(\phi_i, z, t)$. The final threshold value is denoted $q_{ref,i}$, and the corresponding integral is $I(q_{ref,i}, z, t)$. Repeating this process for each element of the array $\phi_i$ yields a one-to-one map $q_{ref}(\phi,z,t)$ that can be approximated to be continuous.

In spherical coordinates, $A$ is therefore expressed as:
\begin{equation}
\mathcal{A}(\lambda, \phi, z, t) \cos\phi 
= -a \int_{\phi}^{\phi(q_{ref})} \big[ q(\lambda, \hat \phi, z, t) - q_{\mathrm{ref}}( \phi, z, t)\big] 
\cos(\hat{\phi}) \, d\hat{\phi},
\label{eq:LWA_def_APP}
\end{equation}
where $a$ is the radius of Earth. The Schematic~\ref{schematic:contoursQref} depicts the integral domain, represented by a blue solid line. We drop the subscript $i$ in Eq.~\ref{eq:LWA_def_APP} as the map $q_{ref}(\phi,z,t)$ is approximated to be continuous. The higher the number of elements of the array $\phi_i$ used to compute $I(q_{\mathrm{ref}})$, the smaller the approximation. Here, we choose a number of array elements twice as large as the resolution of the input dataset. For additional details on equivalent latitude and LWA computation refer to \citet{huang2016local} and \citet{nakamura2010finite}. 

In the present paper, LWA, the associated LWA fluxes, the reference-state $q_{ref}$ and $(u_{ref},v_{ref},\theta_{ref})$ are computed through the \textit{falwa} package (\url{https://github.com/csyhuang/hn2016_falwa}), a Github repository recently published as an open-source tool for LWA computation. For details about the numerical integrations needed for the analysis we redirect the reader to the falwa paper, \citet{huang2025falwa}.
In the rest of the analysis we are not interested in the height dependence of local wave activity dynamics. For the LWA budget equation to be valid we focus on density weighted vertical average of the 3D quantity. Moreover, while the cosine factor explicitly written in eq.~\ref{eq:LWA_def_APP} is important for correctly scaling LWA with latitude, we drop it to improve the readability of the following computations. Hereafter, when referring to LWA, we intend the mass-weighted vertically averaged local wave activity multiplied by the cosine of latitude.  
 
\section{Traffic Jam Theory}
\label{sec:APPENDIX:traffix_jam_theory}

The Traffic Jam theory interprets blocking as a LWA increase resulting from nonlinear interactions between waves and the mean flow. Such non-linear processes can be described with a model analogous to that introduced by \citet{richards1956shock} for a traffic jam on a highway. This analogy becomes quantitative because of the relationship between local wave activity and eddy zonal winds, $u_e$, which follows from the non-acceleration theorem \citep{charney1961propagation}, valid within the WKB condition of a slowly varying medium \citep{huang2016local}.  Note that in this context $u_e$ is the local departure of the zonal wind from the reference-state ($u - u_{ref}$), and it is different from the usual Eulerian definition of an eddy quantity, which is computed as the departure from the climatological temporal mean  ($u - \overline u_{ref}$), and will be hereafter denoted $u'$. 
The relationship is expressed as:
\begin{equation}
u_e = -\alpha A, \qquad \alpha > 0,
\label{eq:uealpha}
\end{equation}
where $u_e$ is the mass-weighted vertical average of eddy wind, $A$ is the local wave activity and $\alpha$ is a positive correlation parameter typically smaller than 1.

The second fundamental assumption required to form the traffic jam analogy concerns the generalized Eliassen-Palm flux relation. Consider its general formulation:
\begin{equation}
\frac{\partial A}{\partial t} = -\nabla \cdot F + \dot{A}
\label{eq:generalized_eliassen_palm_flux}
\end{equation}
where $F$ is the 3D Eliassen-Palm flux and $\dot A$ is the non-conservative sources and sinks of $A$. Assuming that the zonal convergence of LWA dominates the tendency of LWA at synoptic time scales over the meridional or vertical LWA components, one can rewrite Eq.~\ref{eq:generalized_eliassen_palm_flux} as:
\begin{equation}
\frac{\partial A}{\partial t} = -\frac{\partial F_\lambda}{\partial x} + S - \frac{A}{\tau},
\label{eq:lwa_budget}
\end{equation}
with zonal flux $F_\lambda$. The remaining, smaller terms are grouped into $S$, which represents sources and sinks of local wave activity, and $A/\tau$, which accounts for linear damping due to dissipation. This approximation is well supported in the midlatitudes and in storm-track regions, as shown by \citet{barpanda2025local}. The zonal flux  $F_\lambda$ is further decomposed into three contributions:
\begin{equation}
    F_{\lambda} = F_1 + F_2 + F_3,
    \label{eq:APPENDIX:flux_decomposition}
\end{equation}
Here, $F_1$ represents the zonal advective flux of LWA due to the reference-state winds. $F_2$ denotes the zonal nonlinear advective flux of LWA, which is associated with the parameter $\alpha$ introduced in Eq.~\ref{eq:uealpha}, and represents the effect of waves slowing down the background flow. $F_3$ represents the zonal transport of local wave activity associated with wave dispersion. These fluxes can be written explicitly following \citet{nakamura2018atmospheric} or approximated through semi-empirical relations that make the analogy with the Traffic Jam model possible. Both the explicit forms and the approximations are reported in Table~\ref{tab:fluxes}, adapted from \citet{barpanda2025local}.

\begin{table}[htbp]
\centering
\resizebox{\textwidth}{!}{%
\begin{tabular}{llll}
\toprule
Flux & Exact expression & Approx relation with $A$ & Description \\
\midrule
$F_\lambda$ & $F_1 + F_2 + F_3$ & $(u_0 - \alpha A + c_{xg}) A$ & Total zonal flux \\[0.5em]
$F_1$ & $u_{\mathrm{ref}} A$ & $u_0 A$ & Zonal advective flux of LWA due to reference state wind \\[0.5em]
$F_2$ & $- \frac{a}{\cos\phi} \int_\phi^{\phi(q_{ref})} u_e q_e \cos(\hat{\phi}) \, d\hat{\phi}$& $-\alpha A^2$ & Zonal nonlinear advective flux of LWA \\[0.5em]
$F_3$ & $\frac{1}{2}\left(v_e^2 - u_e^2 - \frac{\theta_e^2 \,\mathrm{Re}^{-z/H}}{H\,S_\theta}\right)$ & $c_{xg} A$ & Zonal advective flux of LWA associated with wave dispersion.\\[0.5em]
\bottomrule
\end{tabular}%
}
\caption{Wave activity fluxes and their approximate relations with $  A   $. In these relations, $\theta$ is potential temperature, $v$ is the meridional component of the wind, $H \equiv 7km$ is the assumed scale height, $R$ is the ideal gas constant, $S_\theta$ is the hemispheric-mean static stability given by $\partial_z  \tilde \theta$---where the superscript $\tilde\ $ refers to the hemispheric mean---$c_{xg}$ is the zonal component of the group velocity of Rossby  waves, the variable $u_0$ is time and vertically averaged $u_{ref}$. Moreover, as for Eq.~\ref{eq:uealpha} in the main text, the subscript $(\, \cdot\,)_e$ refers to the anomaly with respect to the reference state. All quantities are density weighted and vertically averaged.}
\label{tab:fluxes}
\end{table}

By making use of the flux approximation, Eq.~\ref{eq:lwa_budget} can thus be rewritten as
\begin{equation}
\frac{\partial A}{\partial t}
= -\frac{\partial}{\partial x} \left[ \left( u_0 + c_{xg} - \alpha A \right) A \right]
+ S - \frac{A}{\tau},
\label{eq:APPENDIX:tjt_eq}
\end{equation}
where $u_0 + c_{xg}$ is hereafter referred to as the Doppler-shifted Rossby-wave group velocity and represents the linear contribution to LWA advection. In this expression, $c_{xg}$ is the zonal component of the Rossby wave group velocity, while $u_0$ is the time and vertically averaged $u_{ref}$. In turn, $-\alpha A$ represents the nonlinear contribution. 

\begin{figure}[htbp]
    \centering
    \includegraphics[width=0.5\textwidth]{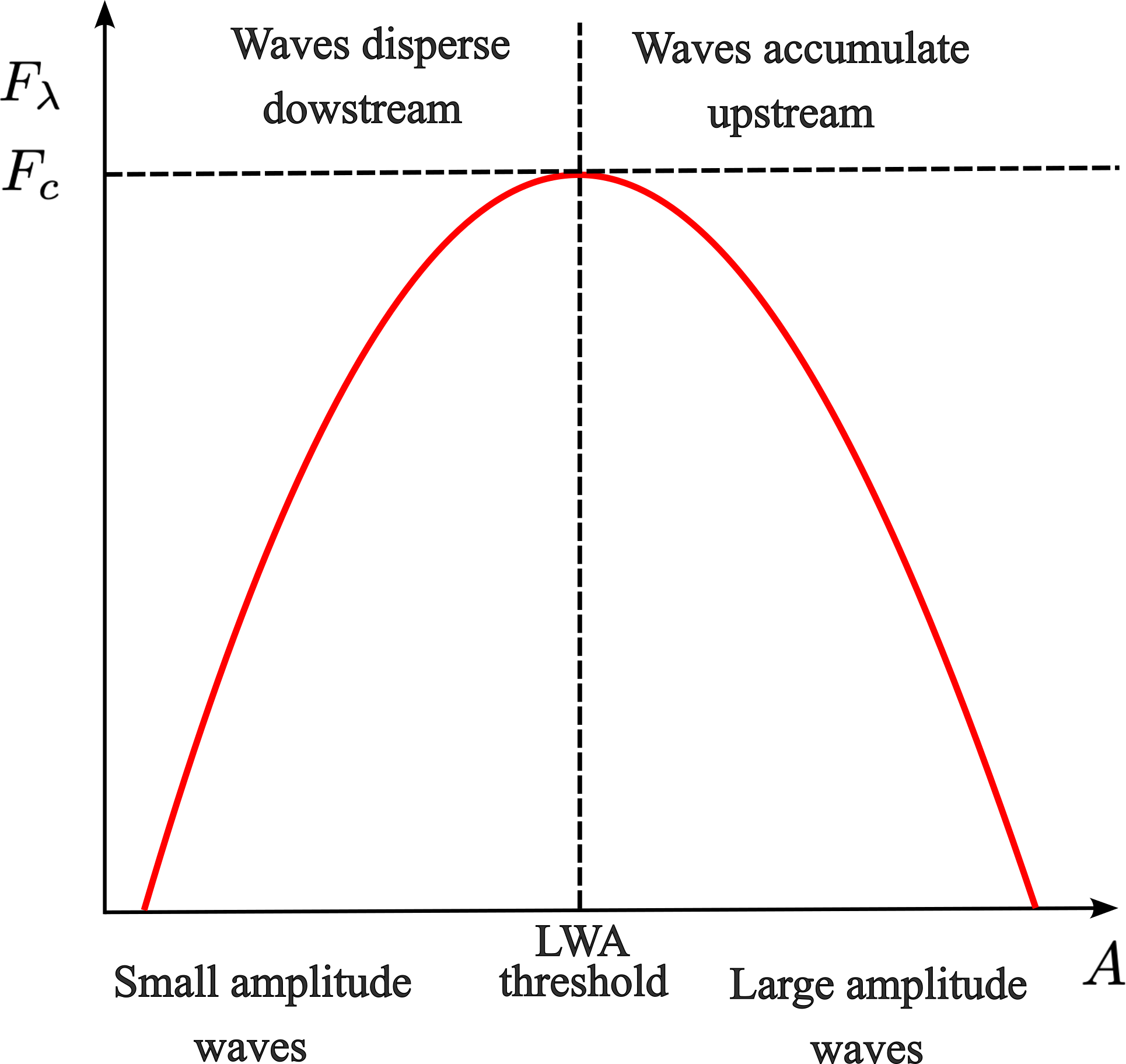}
    \caption{Schematic of the relationship between LWA and the zonal LWA flux. For small-amplitude waves $(LWA\ <<\ threshold)$, the zonal flux increases with LWA due to linear, Doppler-shifted eastward propagation, whereas for large-amplitude waves $(LWA\ >>\ threshold)$ the flux decreases as nonlinear, eddy-induced zonal flow becomes dominant.}
    \label{schematic:parabola}
\end{figure}

The LWA, $A$, in Eq.~\ref{eq:APPENDIX:tjt_eq} is subsequently decomposed into a slowly varying component $A_0$, associated with the stationary Rossby-wave amplitude (discussed in detail in Section~\ref{sec:APPENDIX:carrying_capacity_method}) and a transient eddy component $A'$. Substituting $A(x,t) = A_0(x) + A'(x,t)$ into the budget we obtain a new expression for the transient LWA tendency:
\begin{equation}
\frac{\partial A'}{\partial t}
= -\frac{\partial}{\partial x} 
\underbrace{[(C(x)- \alpha A')A']}_{\text{(i)}} 
+ S - \frac{A'}{\tau},
\label{final_LWA_budget}
\end{equation}
where $C(x)$ is defined as:
\begin{equation}
C(x) \equiv u_0 + c_{xg} - 2\alpha A_0(x).
\label{eq:tjt_eq_updt}
\end{equation}
Term (i) is an expression for the eddy zonal advective flux $F'(A')$ and it is a downward-opening parabola with an absolute maximum, as shown in Fig.~\ref{schematic:parabola} and anticipated in Section \ref{sec:LWA_carrying_capacity_def}. \(F'(A')\) attains a maximum when \(\partial F'/\partial A' = 0\), i.e. at the threshold amplitude
\begin{equation}
A_c(x) = \frac{C(x)}{2\alpha},
\label{eq:Ac_local}
\end{equation}

and the corresponding maximum flux, or carrying capacity $F_c(x)$ defined as
\begin{equation}
F_c(x) = \frac{C(x)^2}{4\alpha}.
\label{eq:Fc_local_APP}
\end{equation}

The carrying capacity $F_c(x)$ provides a useful framework to understand blocking formation. The behavior of the system results from the competition between linear wave propagation and nonlinear wave-mean flow interaction, leading to two distinct dynamical regimes. For $A'(x,t) < A_c$, the linear term $C(x)A'$ dominates the flux $F'(A')$. As the wave amplitude increases, so does the flux. In this state, Rossby waves behave like cars on an open highway: adding more vehicles simply increases the total traffic volume. This corresponds to the normal eastward progression of synoptic storms. 

Once the wave amplitude $A'(x,t)$ grows past the threshold amplitude $A_c(x)$, the nonlinear term $-\alpha A'^2$ becomes dominant. As $A'$ grows, the waves exert a powerful torque that locally decelerates the eddy zonal velocity $u_e$. In this regime, even as the wave amplitude keeps increasing ( $\frac{\partial A'}{\partial t}$ >0), the flux decreases, leading to an upstream accumulation of local wave activity and ultimately to blocking onset. This behavior is analogous to a traffic jam: beyond a critical density, any further increase in the number of vehicles leads to a reduction in the total flow rate due to localized deceleration. The carrying capacity therefore represents the boundary between two different regimes: a "free-flow" regime in which higher LWA is associated with a larger flux and a "congested" regime in which higher LWA opposes the flux and leads to a local accumulation of LWA. These two different regimes are more clearly depicted in Schematic~\ref{schematic:parabola}, adapted from \citet{barpanda2025local}. 

Lastly, we underscore how regions where $C(x)$ is smaller are more prone to blocking formation. These are areas where the stationary Rossby wave amplitude $A_0$ is larger and/or the Doppler-shifted Rossby waves velocity $u_0 + c_{xg}$ is smaller. In this sense, the Traffic Jam theory provides a robust framework to quantitatively estimate the contribution of changes in the background flow to observed or projected changes in atmospheric blocking frequency.

\section{Estimation of the carrying capacity}
\label{sec:APPENDIX:carrying_capacity_method}

In the present analysis, we compare the carrying capacity $F_c$ with the climatological frequency of atmospheric blocking and its response to different forcings. This approach allows us to quantify the individual contribution of changes in the background zonal winds, stationary wave amplitude and the parameter $\alpha$ to changes in blocking frequency. In order to estimate the value of $F_c$ in our experiments we follow the methodology of \citet{barpanda2025local}. We provide a brief overview here and refer interested readers to their work for a comprehensive derivation.

In brief, in order to compute $F_c$ we need to estimate three quantities: the parameter $\alpha$, the Doppler-shifted Rossby waves group velocity $u_0 + c_{xg}$, and the stationary local wave activity $A_0$. The $\alpha$ parameter is derived from Eq.~\ref{eq:uealpha} as the linear regression coefficient between the eddy zonal wind and the local wave activity. This can be expressed as:

\begin{equation}   
\alpha = -\frac{u_eA^T}{A A^T},
\label{eq:alpha_regression}
\end{equation}
where the subscript $^T$ identifies the transpose of the matrix and the other variables have already been defined. The Doppler shifted Rossby wave group velocity is computed through the explicit forms of the zonal LWA fluxes (reported in Table~\ref{tab:fluxes}), which are computed by the \textit{falwa} package. In particular, by combining the zonal advective flux of LWA ($F_1$) and the Rossby wave flux $(F_3)$, we can determine $u_0 + c_{xg}$ as:
\begin{equation}
F_1 + F_3 = (u_0 + c_{gx})A.
\label{eq:flux_regression}
\end{equation}
As for the parameter $\alpha$, the velocity is computed from the slope of the linear regression. For both estimates, the regression is performed point-wise, with the temporal dimension as the free coordinate.

The stationary component $A_0$ is computed as the local wave activity associated with the time averaged QGPV field. This can be expressed mathematically as:

\begin{equation}
\mathcal{A}_0(\lambda, \phi, z, t) \cos\phi 
= -a\, \overline{
\int_{\phi}^{\phi+\phi_C}
\bigl(\overline{q}(\lambda, \hat{\phi}, z, t)
- q_{\mathrm{REF}}(\phi, z, t)\bigr)
\cos(\phi + \hat{\phi}) \, d\hat{\phi}
} \ ,
\label{eq:stationary_lwa_definition}
\end{equation}
where the overbar denotes a temporal average and where we wrote the cosine scaling factor explicitly. 

Prior to these computations  (Eqs.~\ref{eq:alpha_regression},~\ref{eq:flux_regression},~\ref{eq:stationary_lwa_definition}) we apply a Fourier lowpass filter to the LWA field, filtering out variability associated with time scales shorter than 4 days. Moreover, we apply a zonal smoothing a using a 15° moving window. In this way we select only weather systems with blocking timescales and synoptic spatial scales as in \citet{barpanda2025local}.


\noappendix       

\authorcontribution{\textbf{Authors contribution}: MF designed and conducted the research under the supervision of SB and ST. NL assisted MF with the configuration of Isca and with the design of the model experiments. NL, ST and SB contributed to the scientific content of the paper through discussions and numerous meetings with MF. MF wrote the research article and NL, ST and SB contributed to its editing.} 

\competinginterests{\textbf{Competing interests}: The authors declare they have no competing interest} 


\begin{acknowledgements}
\textbf{Acknowledgements}: We thank Dr. Pragallva Barpanda for her help with computing the carrying capacity from our experimental data. We are also grateful to Clare S. Y. Huang, Christopher Polster, and Noboru Nakamura, the authors of the falwa package, for making their software freely available and allowing us to use it in our analysis.
This research was conducted by MF within the Italian national inter-university doctoral program \emph{Sustainable Development and Climate Change} (\href{https://www.phd-sdc.it/}{https://www.phd-sdc.it}). SB acknowledges support from the National Recovery and Resilience Plan (NRRP), Mission 4, Component 2, Investment 1.4 (Call for Tender No.~1031 of 17/06/2022) of the Italian Ministry for University and Research, funded by the European Union--NextGenerationEU (Project No.~CN\_00000013) and the Italian Ministry of University and Research in the framework of the PRIN 2022 call - Protocol no. 2022WT939B, CUP E53C24002980006.
\end{acknowledgements}

\codeavailability{\textbf{Code availability}: The scripts used for the manuscript analysis are available at \href{https://github.com/michele-filippucci/lwa_persistence}{github.com/michele-filippucci/lwa\_persistence}. Isca is available as an open source software at \href{https://github.com/ExeClim/Isca}{github.com/ExeClim/Isca}, while the falwa package is available at \href{https://github.com/csyhuang/hn2016_falwa/}{github.com/csyhuang/hn2016\_falwa/}.}





\bibliographystyle{copernicus}
\bibliography{export.bib}

\end{document}


\maketitle

In this supplementary material we analyze the mechanisms responsible for the decrease of $\alpha$, the locally-defined negative correlation field between LWA and the eddy zonal winds. The $\alpha$ field can be computed from the eddy zonal wind field ($u_e$) and the density weighted vertically averaged Local Wave Activity (LWA or $A$) through the relation 
\begin{equation}
\alpha = - \frac{u _eA^{T}}{A A^{T}},
\end{equation}

We can therefore obtain additional insights on the $\alpha$ decrease by decomposing its response to the Arctic forcing into two distinct contributions: (i) changes in the variance of LWA, $A A^{T}$ (hereafter just `variance') , and (ii) changes in the covariance between LWA and the eddy zonal wind, $u _eA^{T}$ (hereafter just `covariance'). Writing
\[
\alpha = - \frac{C}{V},
\quad
\text{with}
\quad
C \equiv u_eA^{T}, 
\quad
V \equiv A A^{T},
\]
the variation of $\alpha$ can be rewritten as:
\begin{equation}
\Delta \alpha = -(\Delta C / V - (C/V^{2})\Delta V) = - (C/V) (\Delta C / C - \Delta V / V) = - \alpha (\Delta C / C - \Delta V / V).
\label{eq:alpha_decomposition}
\end{equation}

This decomposition allows us to isolate whether the weakening of $\alpha$ is primarily driven by modifications in the amplitude of LWA fluctuations or by changes in their dynamical coupling with the mean flow. The analysis is shown for the SYM simulations, but we obtain the same results for the ASYM configuration.

\begin{figure}[htbp]
    \centering
    \includegraphics[width=1\textwidth]{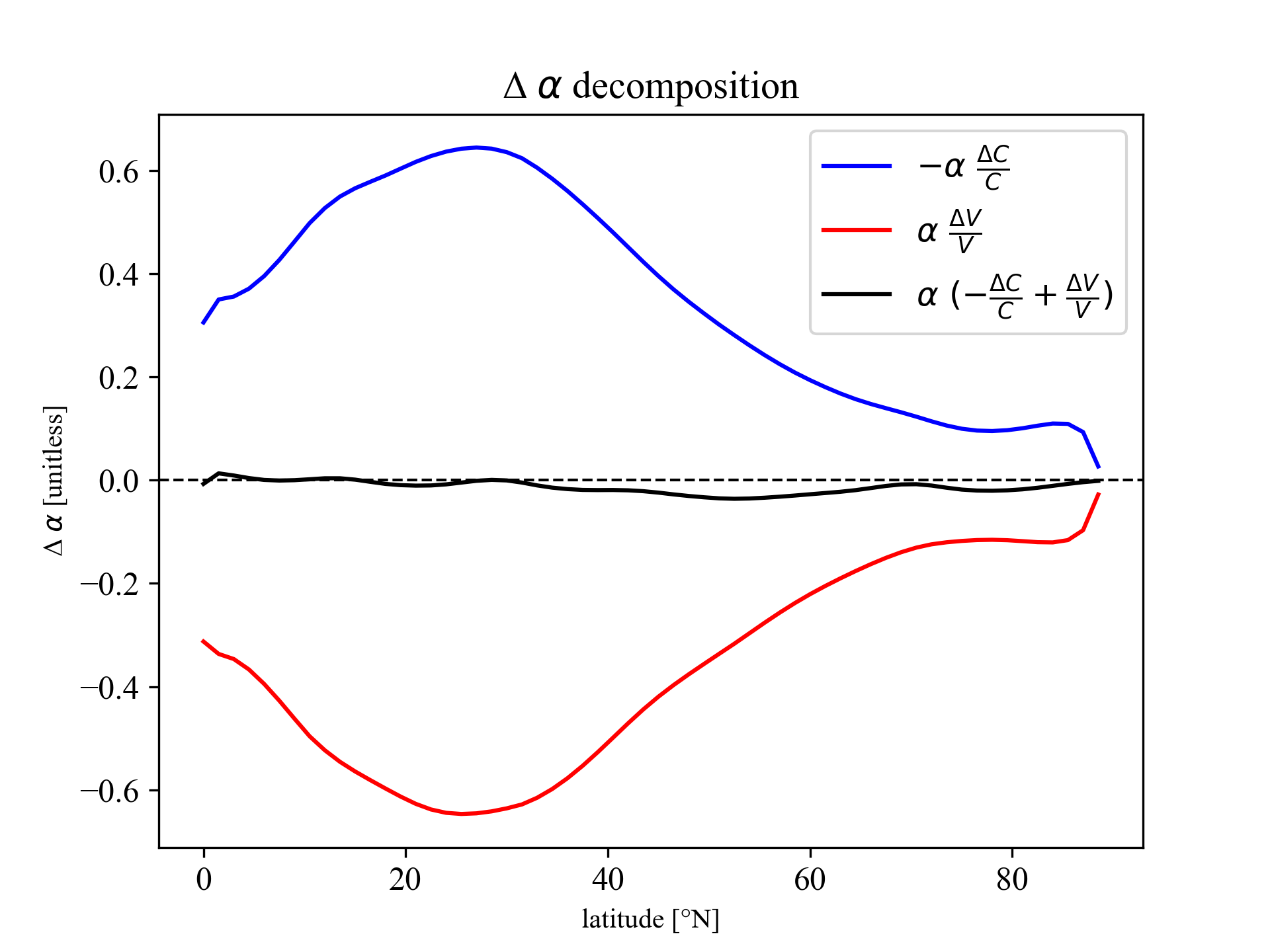}
    \caption{Decomposition of changes of $\alpha$  (black line) into the contribution of LWA and eddy zonal wind covariance changes ($-\alpha \frac{\Delta C}{C}$) (blue line) and LWA variance changes  ($\alpha \frac{\Delta V}{V}$) (red line). Changes are zonally averaged.}
    \label{fig:alpha_response}
\end{figure}

Fig. \ref{fig:alpha_response} shows the decomposition terms described in Eq. \ref{eq:alpha_decomposition}. At midlatitudes, the increase of variance and changes of covariance act in the same direction (i.e. both quantities are negative), leading to a decrease of $\alpha$. On the contrary, at high latitudes the reduction of variance and the strong decrease of covariance act in opposite directions, resulting in a small decrease of $\alpha$. The same happens at low latitudes.

\begin{figure}[htbp]
    \centering
    \includegraphics[width=1\textwidth]{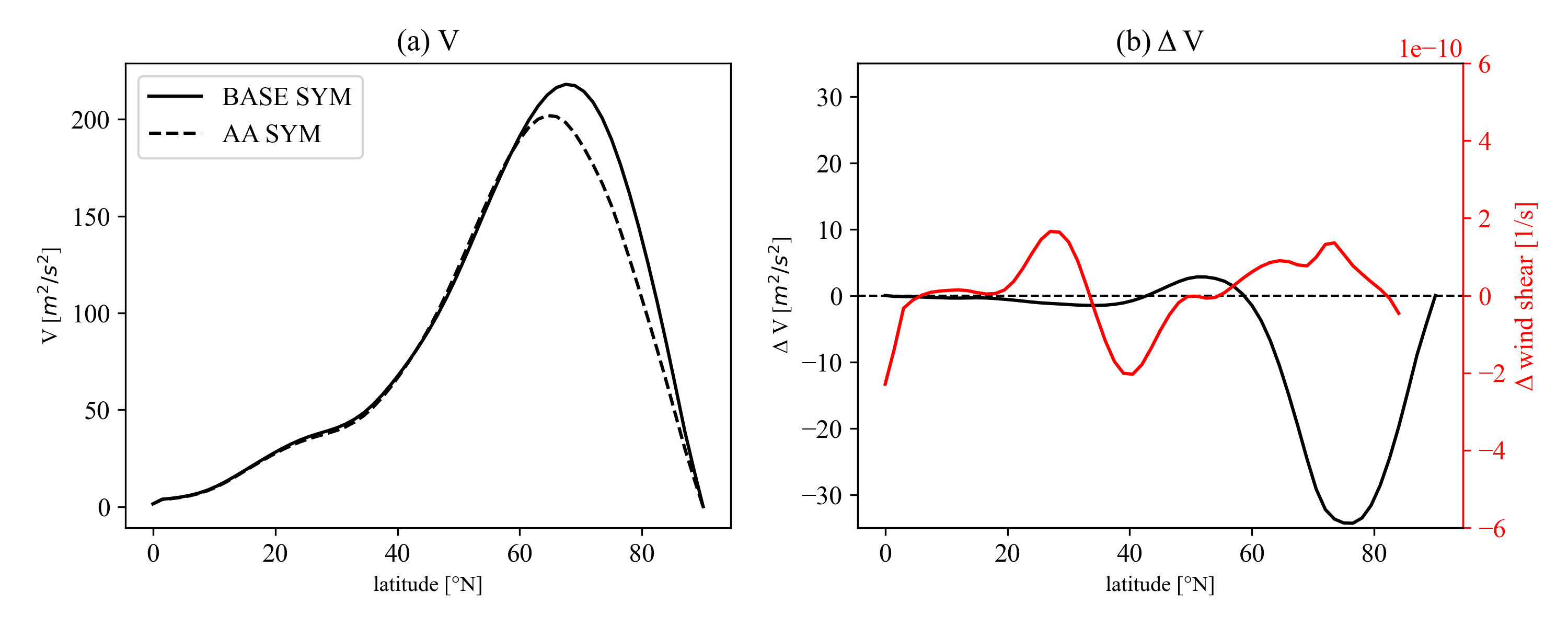}
    \caption{Panel a shows the zonally averaged LWA variance [$m^2/s^2$] for both the BASE SYM (dashed line) and the AA SYM (solid) simulations. Panel b compares the change in LWA variance (AA-BASE) (black line and axes) with changes in meridional wind shear [$1/s$] (red line and axes).}
    \label{fig:alpha_decomposition_variance}
\end{figure}

To interpret the physical meaning of these variations we analyze separately changes of variance and covariance. Starting with the covariance (Fig.~\ref{fig:alpha_decomposition_variance}) we find that it decreases at high latitudes, while it slightly increases at mid latitudes. We propose an interpretation of these differences based on changes of the meridional wind shear. Earlier literature highlighted how the jet stream can act as a wave guide: waves propagate preferentially in regions of low wind shear, near the jet core, while in regions of strong shear, poleward and equatorward of the jet, they tend to break because of the wave straining mechanism and the growing importance of non-linear terms of the vorticity equation \citep{andrews1976,mcintyre1983}. We therefore expect changes of wind shear to affect the variance, which depends strongly on the presence and propagation of synoptic eddies. Fig. \ref{fig:alpha_decomposition_variance}b shows changes in both wind shear and variance, highlighting how they are negatively correlated. An offset exists between the two, which may be linked to the general decrease of baroclinicity from mid to high latitudes, evident by looking at isothermals in Figure 4 in the main text, or to the overall wind speed decrease at mid-to-high latitudes that decreases the zonal flux of LWA.

\begin{figure}[htbp]
    \centering
    \includegraphics[width=1\textwidth]{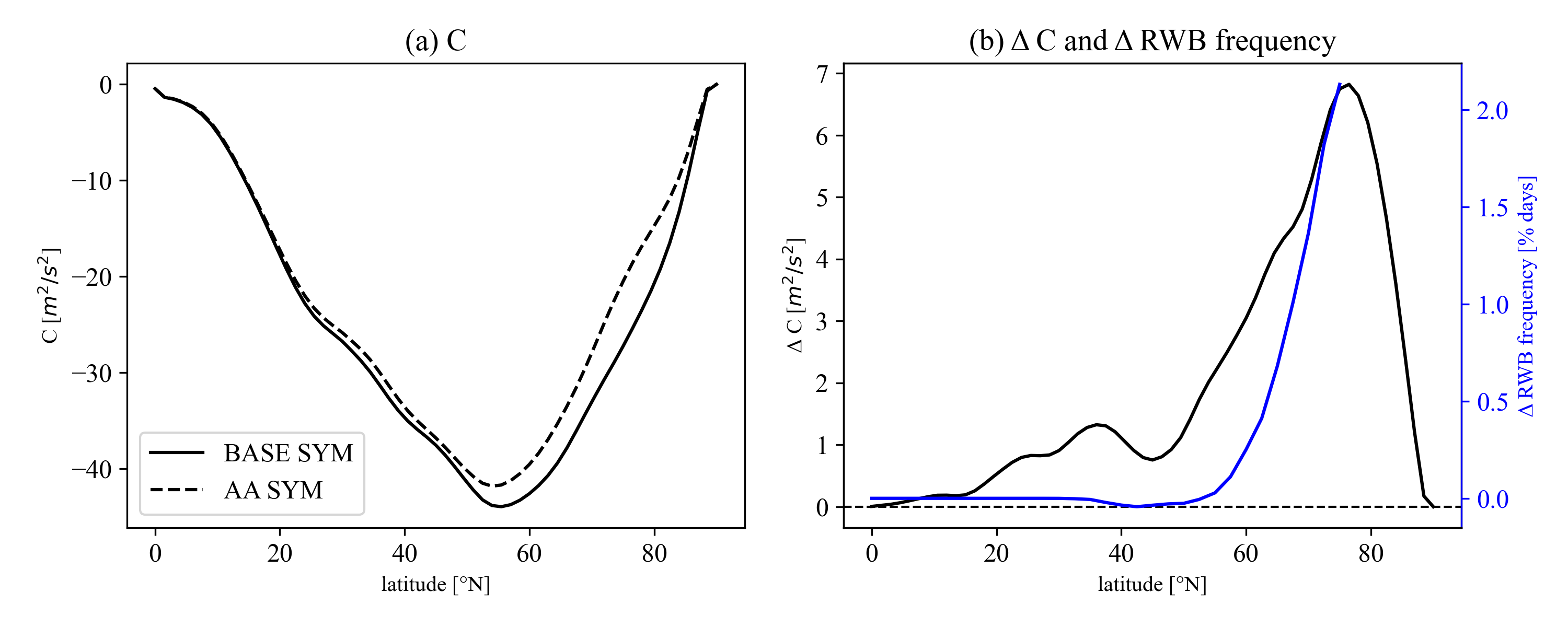}
    \caption{Panel a shows the zonally averaged LWA covariance [$m^2/s^2$] for both the BASE SYM (dashed line) and the AA SYM (solid) simulations. Panel b compares the change in LWA covariance (AA-BASE) (black line and axes) with changes in RWB frequency [$\%\  days$] (blue line and axes).}
    \label{fig:alpha_decomposition_covariance}
\end{figure}

Moving to changes of covariance (Figure \ref{fig:alpha_decomposition_covariance}), we find differences concentrated at high latitudes, where a substantial increase is observed. We speculate that this increase is again linked to wind shear differences, which act to increase the frequency of Rossby wave breaking (RWB) at high latitudes. As described in the main text, when the LWA flux surpasses the maximum carrying capacity -- which at high latitudes is rather low -- RWB occurs and zonal winds are deflected, while LWA keeps accumulating. In this regime, 
the relationship between zonal winds and LWA changes and the covariance becomes weaker. Note that a weakened negative covariance implies an overall covariance increase.  Figure \ref{fig:alpha_decomposition_covariance}b compares changes of covariance with changes of RWB frequency under Arctic warming, computed thanks to the Davini index \citep{davini2012bidimensional}, which looks for the reversal of the meridional gradient of geopotential height at 500 hPa (refer to \citet{davini2012bidimensional} for additional details). We underscore that RWB events detected by the Davini index are quite different from the blocking events detected through our method based on LWA anomalies, adopted in the main text. The reversal of the geopotential height gradient does not necessarily lead to LWA anomalies large enough to be detected as blocking events by our method. In fact, events detected by the Davini index are concentrated at high latitudes and their frequency distribution is considerably different from that of our LWA anomaly-based index, with almost no overlaps (not shown). As already reported by \citet{ronalds2019role}, RWB at highlatitudes increases following Arctic warming. Changes in RWB frequency and covariance are therefore negatively correlated. While these results support our interpretation, they do not prove a causal relationship between changes of RWB frequency and LWA and wind covariance. A more thorough assessment of causal relationships is left for future work.

Returning to the interpretation of Fig.~\ref{fig:alpha_response}, we found that changes in wind shear and RWB compensate each other in high-latitudes, while in midlatitudes the narrowing of the midlatitude wave-guide acquires a greater relative importance. Summarizing, we conclude that changes in the meridional shear of zonal winds lead to a small but significant decrease of $\alpha$, slightly increasing the carrying capacity of the flow, as shown in the main text. 

\bibliographystyle{copernicus}  
\bibliography{supplementary_references}  